\documentclass{jfm}
\usepackage{graphicx}
\usepackage{epstopdf, epsfig}
\usepackage{amsmath}
\usepackage{commath}
\DeclareMathOperator{\sign}{sgn}
\usepackage[section]{placeins}
\usepackage{xcolor}
\usepackage{float}

\shorttitle{Motion of a sphere in a stratified fluid}
\shortauthor{Arun Kumar V. and G. Subramanian}
\title{Motion of a sphere in a viscous density stratified fluid}
\author{Arun Kumar Varanasi,
  Ganesh Subramanian\corresp{\email{sganesh@jncasr.ac.in}}}
\affiliation{Engineering Mechanics Unit, Jawaharlal Nehru Center for Advanced Scientific Research, Bangalore-560064, India}

\begin{document}
\maketitle
\begin{abstract}
We examine the translation of a sphere in a stably stratified ambient in the limit of small Reynolds ($Re \ll 1$) and viscous Richardson numbers ($Ri_v \ll 1$); here, $Re = \frac{\rho Ua}{\mu}$ and $Ri_v = \frac{\gamma a^3 g}{\mu U}$ with $a$ being the sphere radius, $U$ the translation speed, $\rho$ and $\mu$ the density and viscosity of the stratified ambient, $g$ the acceleration due to gravity, and $\gamma$ the density gradient (assumed constant) characterizing the ambient stratification. In contrast to most earlier efforts, our study considers the convection dominant limit corresponding to $Pe = \frac{Ua}{D} \gg 1$, $D$ being the diffusivity of the stratifying agent. We characterize in detail the velocity and density fields around the particle in what we term the Stokes stratification regime, defined by $Re \ll Ri_v^{\frac{1}{3}} \ll 1$, and corresponding to the dominance of buoyancy over inertial forces. Buoyancy forces associated with the perturbed stratification fundamentally alter the viscously dominated fluid motion at large distances. At distances of order the stratification screening length, that scales as $aRi_v^{-\frac{1}{3}}$, the motion transforms from the familiar fore-aft symmetric Stokesian form to a fore-aft asymmetric pattern of recirculating cells with primarily horizontal motion within; except in the vicinity of the rear stagnation streamline. At larger distances, the motion is vanishingly small except within (a) an axisymmetric horizontal wake whose vertical extent grows as $O(r_t^{\frac{2}{5}})$, $r_t$ being the distance in the plane perpendicular to translation and (b) a buoyant reverse jet behind the particle that narrows as the inverse square root of distance downstream. As a result, for $Pe = \infty$, the motion close to the rear stagnation streamline starts off pointing in the direction of translation, in the inner Stokesian region, and decaying as the inverse of the downstream distance; the motion reverses beyond a distance of $1.15aRi_v^{-\frac{1}{3}}$, with  the eventual reverse flow in the far-field buoyant jet again decaying as the inverse of the  distance downstream. For large but finite $Pe$, the narrowing jet is smeared out beyond a distance of $O(a Ri_v^{-\frac{1}{6}} Pe^{\frac{1}{2}})$, leading to an exponential decay in the aforementioned reverse flow.
\end{abstract}

\section{Introduction}\label{intro}
The phenomena of particles moving in a density stratified environment is a common occurrence in nature since both the atmosphere and the oceans are, on average, stably stratified. Considering the oceans, for instance, there exist examples of both active (aquatic swimming organisms) and passive (so-called marine snow) particles moving through the stratified pycnocline \citep{magnaudet_particles_2020}, the former often as part of a diurnal migration pattern that has been termed the largest migration on earth \citep{martin_2020}. This work was originally motivated by a rather provocative suggestion \citep{katija_2009,subramanian_2010} of the aforementioned migratory pattern leading to an additional biogenic contribution to the mixing of the ocean waters; this, in addition to the two well known mechanisms of winds and tides \citep{munk_1966}. In contrast to the latter two, the energy input in the proposed biogenic contribution occurs at the smallest scales, since the vast majority of the aquatic biomass is concentrated at these scales (the zooplankton or copepods involved in the migration range in size from tens of microns to a few millimeters) \citep{kunze_2006,visser_2007}. As evident from the arguments put forth in \cite{katija_2009}, the validity of the biogenic mixing hypothesis is rooted in the ability of a single small active organism, or a passive particle, to drag along a large amount of fluid during its migration, thereby contributing to the (vertical) mixing of the ocean waters on larger scales. Interestingly, in a homogeneous fluid medium and for any finite Reynolds number, a passive particle can drag an arbitrarily large volume of fluid, over sufficiently long times, on account of the slowly decaying velocity field in its viscous wake \citep{eames_2003,chisholm_2017}. However, as pointed out by \cite{subramanian_2010}, the oceans being stably stratified, this dragging motion incurs a potential energy penalty on large enough length scales. The limit of a vanishing stratification (corresponding to a homogeneous fluid medium) is therefore a singular one; in that, a small but finite stratification is expected to render the volume dragged by the particle, the so-called drift volume \citep{darwin_note_1953,lighthill_1956}, finite.

The above description makes it clear that, at the heart of the validity of the biogenic mixing hypothesis, is the nature of fluid motion induced by an active or passive particle in a stably stratified medium. This study examines the latter problem, that of a small passive particle translating in a stably stratified medium, where `small' refers to the dominance of viscous forces. Consideration of a passive particle is not overly restrictive since even active swimmers, moving along the vertical, attain neutral buoyancy only at a certain instant in time (corresponding to a depth at which the ambient and swimmer densities equal each other). At all other times, such swimmers exert a net force on the ambient. Despite the near-field being dominated by the fluid motion induced by the slip velocity on the swimmer surface \citep{doostmohammadi_2012}, one expects the net force to invariably play a dominant role in the far-field. With this in mind, we consider a passive sphere translating along the direction of stratification at small Reynolds($Re$) and viscous Richardson numbers($Ri_v$), the translation assumed to be the result of a density difference. $Ri_v =\frac{\gamma a^3 g}{\mu U}$ measures the relative importance of viscous and buoyancy forces, and is therefore the key dimensionless parameter for motion of small particles in a stratified ambient; here, $\gamma=-\frac{d\rho}{dz}$ is the constant density gradient in the ambient($\gamma>0$ for stable stratification), $a$ the sphere radius, $g$ the acceleration due to gravity, $\mu$ the fluid viscosity and $U$ the speed of translation. Note that $Ri_v=\frac{Re}{Fr^2}$, where $Fr= \frac{U}{Na}$ is the Froude number that is the usual measure of the importance of stratification in the inviscid limit, $N=\sqrt{-\frac{g\gamma}{\rho}} $ here being the Brunt-Vaisala frequency \citep{turner_1979}. In a significant departure from most earlier efforts (discussed below), and keeping in mind the oceanic scenario, we consider the Peclet number, defined as $Pe = \frac{Ua}{D}$, $D$ being the diffusivity of the stratifying agent (salt in the oceanic case) to be large. 

As mentioned above, earlier efforts, particularly the ones devoted to analysis of the fluid motion around a moving particle or swimmer, have mostly been restricted to small $Pe$; an exception is the very recent effort of \citet{shaik_2020b}, and we discuss this in section  \ref{conclusions}. Motivated by the need to understand laminar jets in a stratified ambient, \cite{list_laminar_1971} was the first to characterize the analog of a Stokeslet (the limiting scenario of a small translating particle, for $Re = 0$, approximated as a point force) in a linearly stratified fluid, and for small $Pe$. The author considered both vertical and horizontal Stokeslet orientations in two and three dimensions; for the vertical orientation, relevant to the problem analyzed here, the motion although fore-aft symmetric was shown to decay away much more rapidly than the $O(\frac{1}{r})$ decay characteristic of a Stokeslet in a three-dimensional homogeneous ambient. The resulting weak far-field motion, shown in the paper only for the two-dimensional case, was in the form of toroidal recirculation cells stacked along the direction of stratification. `Far-field' here refers to (in units of $a$) length scales of $O(Ri_vPe)^{-\frac{1}{4}}$, the stratification screening length for $Pe \ll 1$; as will be seen below, the number of such cells is finite. Much later, \citet{ardekani_2010} considered the same problem, but for both passive and active particles modeled as point force and force-dipole singularities, respectively. The density and velocity fields were obtained numerically using a fast Fourier transform technique, the singularities being termed `stratlets'. More recently, \citet{fouxon_2014} examined the role of turbulence, within the Boussinesq framework, in disrupting the stratification-induced signatures on the flow field around passive particles and active swimmers. As part of their analysis, the authors derived an asymptotic expression for the far-field flow in the absence of turbulence, and that exhibited a rapid algebraic decay, consistent with the findings of the aforementioned studies. \citet{wagner_2014} examined the mixing efficiencies associated with the flow induced by micro-swimmers, for small $Pe$, finding them to be negligibly small. Very recently, \citet{mercier_2020} and \citet{dandekar_2020} have analyzed the drag and torque acting on anisotropic disk-shaped particles (and the resulting orientation dynamics) sedimenting in a stratified medium. The experiments reported in \citet{mercier_2020} pertain to finite $Re$ and $Ri_v$, and highlight the existence of an edgewise-settling regime for sufficiently large $Ri_v$ or small $Fr$ (in this regard, also see \cite{ardekani2014}; \citet{mrokowska_2018,mrokowska_2020a,mrokowska_2020b}); in contrast to the broadside-on settling regime known for small to moderate Re in a homogeneous ambient \citep{cox_1965,dabade_2015,prateek_2020}. The theoretical effort of \citet{dandekar_2020} evaluates the hydrodynamic force and torque on an arbitrarily shaped body in a linearly stratified ambient for arbitrary $Pe$, and finds a hydrodynamic torque, arising from the ambient stratification, for chiral particles. The role of stratification in the orientation dynamics of achiral particles, such as the ones used in \citet{mercier_2020}, has been analyzed in \citet{stratified_torque_2021}. In the present context, we only note that, although the aforementioned recent studies also pertain to the large-$Pe$ limit, the fluid motion was not examined in detail.

As seen above, a number of efforts in the literature have analyzed the fluid motion around both passive particles and active swimmers primarily in the small $Pe$ regime. However, the motion of a typical particle or small-sized swimmer (zooplankton) in the oceanic ambient, relevant to the biogenic mixing hypothesis, pertains to large $Pe$; for instance, a zooplankton of size $0.1$ $mm$ moving at a speed of $1$ $mm/s$ in a typical oceanic stratification of  $\gamma=1.67\times10^{-3} \frac{kg}{m^4}$, yields $Re=0.116$, $Ri_v=1.84\times10^{-8}$ and $Pe=100$. Note that the large $Pe$ regime pertains generically to cases where salt is the stratifying agent, for particles larger than a few microns, the aforementioned oceanic ambient only being one such instance. The first theoretical effort in this regime is that of \citet{zvirin_settling_1975} who calculated the drag enhancement in what we term the Stokes stratification regime below, and is defined by $Re \ll Ri_v^{\frac{1}{3}} \ll 1$. The calculation was restricted to determining the drag enhancement arising from buoyancy effects in the outer region, on scales of $O(Ri_v^{-\frac{1}{3}})$, corresponding to the stratification screening length (note that this is the screening length for large $Pe$, in contrast to the $O(Ri_v Pe)^{-\frac{1}{4}}$ screening length above, for small $Pe$, that was obtained by \citet{list_laminar_1971} and \citet{ardekani_2010}). Similar to Childress's determination of the drag correction for the axial motion of a sphere in a rotating fluid\citep{childress_1964}, and Saffman's calculation of the inertial lift \citep{saffman_1965}, the analysis was done in Fourier space, with the correction to the Stokes drag coming out to be $O(Ri_v^{\frac{1}{3}})$, the inverse of the aforementioned screening length. More recently, \citet{zhang_core_2019}, by using detailed numerical calculations and an ingenious splitting procedure, showed that the enhancement in drag at low Reynolds numbers comes from the induced baroclinic torque and the resulting change in the flow structure. Moreover, the enhancement in drag was found to be proportional to $Ri_v^{\frac{1}{3}}$, in agreement with the theoretical result above. These results, however, do not agree with the observations of \citet{yick_2009} who obtained a scaling closer to $Ri_v^{\frac{1}{2}}$, the mismatch likely due to additional non-Boussinesq contributions arising from heavily deformed iso-pycnals close to the sphere. A recent effort of \cite{mehaddi_2018} has extended the sphere drag calculation to include effects of weak inertia.

The primary motivation for our calculation is to eventually determine the drift volume in a stably stratified ambient, and thereby, estimate the importance of the biogenic mixing contribution. Now, as mentioned above, the infinite-time drift volume is divergent, for any finite Re, in a homogeneous ambient \citep{eames_2003,chisholm_2017,subramanian_2010}, this divergence arising from the slow $O(\frac{1}{r})$ decay of velocity field within the far-field wake, $r$ being the distance downstream. For $Re = 0$, the velocity field decays as $O(\frac{1}{r})$ at large distances regardless of the direction, and as a result, the drift volume diverges for any finite time. This implies that the finiteness of the drift volume, for a weakly stratified ambient pertaining to the aforementioned Stokes stratification regime, must arise from the transition of the far-field fluid motion from an $O(\frac{1}{r})$ Stokesian decay to a more rapid decay beyond the $O(Ri_v^{-\frac{1}{3}})$ stratification screening length. Thus, for small $Re$, and unlike the drag problem considered in \citet{zvirin_settling_1975}, one expects the dominant contribution to the drift volume to arise from the fluid motion far from the sphere, or in other words, the outer region. It is with this in mind that the analysis here is restricted to the linearized equations in the far-field. One may nevertheless question the relevance of this linearization, given that the motion in the outer region is indirectly influenced by the heavily deformed iso-pycnals, close to the sphere, for large $Pe$. However, these deformed iso-pycnals contribute to a localized buoyant envelope around the sphere, and at large distances, one may regard the combination of the envelope and the sphere as an effective point force, albeit of a different magnitude, as far as the outer region is concerned; the linearity of the outer-region equations implies that the nature of fluid motion is independent of the magnitude of the force. More detailed scaling arguments pertaining to the velocity and density fields in the inner region (length scales of order the particle size) are given in the conclusions section.

The remainder of the paper is organized as follows. In the next section, we present the quasi-steady governing equations for the fluid motion under the Boussinesq approximation and a scaling analysis to determine the screening lengths arising from the effects of inertia and stratification, for both small and large $Pe$. Next, the linearized equations in the outer region are solved using a Fourier transform approach \citep{saffman_1965,childress_1964}, and the velocity and density field are written as Fourier integrals, in the aforementioned small and large-Pe limits, and in the so-called Stokes stratification regime, when buoyancy forces are dominant over inertial ones; this translates to $Re \ll (Ri_vPe)^{1/4}$ for small $Pe$, and $Re \ll Ri_v^{1/3}$ for large $Pe$. In section 3, we contrast the streamline patterns and iso-pycnals obtained from a numerical evaluation of the Fourier integrals for $Pe = 0$ and $Pe \gg 1$; the numerical results are also compared to analytical approximations valid for distances much greater than the respective screening lengths. In the concluding section \ref{conclusions}, we summarize our work, and follow this up with scaling arguments pertaining to the inner region dynamics and drift volume. 

\section{The disturbance fields in a stable linearly stratified ambient}\label{goveq}
We consider a sphere of radius $a$ moving vertically with speed $U$ in an unbounded stably stratified fluid with a linear stratification profile $\frac{d\rho}{dz}=-\gamma$, with $\gamma > 0$. Using $a$, $U$ and $\gamma a$ for the length, velocity and density scales, respectively, the non-dimensional continuity equation, the Navier-Stokes equations and the convection-diffusion equation for the velocity($\mathbf{u}$) and density disturbance($\rho_f$) fields, in a sphere-fixed reference frame, are as follows:

\begin{equation}
\nabla \cdot \mathbf u=0,
\end{equation}
\begin{equation}
Re[\mathbf u \cdot \nabla \mathbf u]=-\nabla p+\nabla^2 \mathbf u - Ri_v \rho_f \mathbf{1_z},
\end{equation}
\begin{equation}\label{eq:dens}
1-w+\mathbf u \cdot \mathbf{\nabla} \rho_f= \frac{1}{Pe} \nabla^2 \rho_f, 
\end{equation}

\begin{align}
\mathbf{u}=0, \quad \mathbf{n} \cdot \nabla{\rho_f} =0 \quad \mbox{ at } \quad r=|\mathbf{x}|=1, \\
\mathbf{u} \rightarrow \mathbf{1_z}, \quad  \rho_f \rightarrow 0 \quad \mbox{ as } \quad r=|\mathbf{x}| \rightarrow \infty,
\end{align}
where $r$ is the non-dimensional distance from the sphere and $w$ in (\ref{eq:dens}) is the vertical velocity component. The total density in the aforementioned reference frame is given by  $\rho(z)=\rho_0+t-z+\rho_f$,  and the term involving $1-w$ in (\ref{eq:dens}) denotes the convection of the base-state stratification\,(along the vertical coordinate) by the perturbation velocity field. Note that the Boussinesq approximation has been used above to neglect the density disturbance in the convective terms of the equations of motion, so  $Re$ in $(2.2)$ is based on an appropriate reference density. Further, in taking $\rho_f$ in particular to be independent of time, we have assumed a quasi-steady state to be achieved for long times. This assumption is examined in section $4$ for both the inner ($r \sim O(a)$) and outer regions ($r \geq O(Ri_v^{-\frac{1}{3}})$).

As is well known, although we examine the limit $Re, Ri_v \ll 1$, the inertial and stratification terms in ($2.2$) cannot be neglected. This is because the resulting Stokes equations are not a uniformly valid approximation, and the aforementioned terms become comparable to the leading order viscous terms at sufficiently large distances. As discussed in the introduction, the large length scales above are precisely the ones that control the drift volume that in turn underlies the biogenic mixing hypothesis. For a homogeneous fluid, the length scale (in units of $a$) at which inertial forces first become comparable to viscous forces is $Re^{-1}$, referred to here as the inertial screening length. Obtaining a similar estimate for the buoyancy forces requires one to obtain the far-field behavior of the density field which in turn depends on whether $Pe$ is large or small.

For $Pe \rightarrow 0$, the density perturbation on length scales of $O(a)$ arises from the no-flux boundary condition on the surface of the particle, and decays as $O(\frac{1}{r^2})$ at large distances. The convective correction to the density field satisfies $\frac{1}{Pe} \nabla^2 \rho_f \sim (1-w)$; using $(1-w) \sim O(\frac{1}{r})$ for the Stokeslet field leads to $\rho_f \sim Pe \; r$. The buoyancy forces arising from the convective perturbation are $O(Ri_v Pe\:r)$, and grow linearly with distance. Equating them to the decaying viscous forces of $O(\frac{1}{r^3})$ leads to the small-$Pe$ stratification screening length $l_c \sim  (Ri_vPe)^{-\frac{1}{4}}$. The equations governing the disturbance fields on scales of order the aforementioned screening length may be obtained by using the expansions: $\mathbf{u}=\mathbf{1_z}+(Ri_vPe)^{1/4}\mathbf{\bar u}$, $p=p_{\infty}+(Ri_vPe)^{\frac{1}{2}}\bar{p}$ and $\rho_f=Pe(Ri_vPe)^{-\frac{1}{4}}\bar{\rho_f}$. Note that the velocity, pressure and density disturbance vary as $\frac{1}{r}$, $\frac{1}{r^2}$ and $Pe\:r$, respectively, in the inner Stokesian region far away from the particle, leading to the scalings in the above expansions. The outer region equations for $\mathbf{\bar u}$, $\bar p$ and $\bar \rho_f$ are given by

\begin{equation}\label{eq:cont_Pe0}
\bar \nabla. \mathbf{\bar u}=0,
\end{equation}
\begin{equation}\label{eq:vel}
-\alpha_0 \frac{\partial \mathbf{\bar u}}{\partial \bar z}= -\bar \nabla \bar p +\bar \nabla^2 \mathbf{\bar u}-[\bar {\rho_f}+6\pi\delta(\mathbf{ \bar r})]\mathbf{1_z},
\end{equation}
\begin{equation}\label{eq:dens_Pe0}
-\mathbf{1_z}.\bar u+\beta_0 \frac{\partial\bar{\rho_f}}{\partial \bar z} = \bar \nabla^2 \bar{\rho_f}.
\end{equation}
Here, $\alpha_0$ and $\beta_0$ are given by $\frac{Re}{(Ri_vPe)^{1/4}}$ and $\frac{Pe}{(Ri_vPe)^{1/4}}$, respectively, and denote the ratios of the low-$Pe$ stratification screening length to the inertial ($Re^{-1}$) and convective($Pe^{-1}$) screening lengths. Note that the boundary condition on the particle surface has now been replaced by a point force on the RHS of (\ref{eq:vel}). For $Pe,Re \ll (Ri_vPe)^{\frac{1}{4}}$, one may ignore the terms proportional to $\alpha_0$ and $\beta_0$. One then finds the velocity and density disturbance fields as the following Fourier integrals:
\begin{equation}\label{eqn:axvellow}
\mathbf{\bar{u}} (\mathbf{\bar{r}}) = \frac{-3}{4\pi^2} \int \frac{k^4 (\mathbf{1_z}-\frac{k_3\mathbf{k}}{k^2})}{k^6+k_t^2} e^{i\mathbf k.\mathbf{\bar r}} d\mathbf{k},
\end{equation}
\begin{equation}\label{eqn:densdislow}
{\bar{\rho_f}}(\mathbf{\bar{r}}) = \frac{-3}{4\pi^2} \int \frac{k_t^2}{k^6+k_t^2} e^{i\mathbf k.\mathbf{\bar r}} d\mathbf{k},
\end{equation}
where $k_t = (k^2-{k_3}^2)^{\frac{1}{2}}$ is the magnitude of the wavevector projected onto the plane perpendicular to the translation direction. The above diffusion dominant limit has been considered previously\,(see \citep{list_laminar_1971,ardekani_2010,fouxon_2014}), as indicated in the introduction, and we have included this case only for purposes of contrasting with the results obtained below in the convection dominant limit.

For $Pe \rightarrow \infty$, one neglects the diffusion term in ($2.3$) and thus $\mathbf{u} \cdot \nabla\rho_f \sim (1-w)$. Again, using $(1-w) \sim O(\frac{1}{r})$, one has $\rho_f \sim O(1)$, so the buoyancy forcing term in ($2.2$) is $O(Ri_v)$. Equating this to the $O(l_c^{-3})$ viscous term, one obtains the large-$Pe$ stratification screening length to be $l_c \sim Ri_v^{-\frac{1}{3}}$, as originally shown by \citet{zvirin_settling_1975}. Again, keeping in mind the Stokesian scalings in the inner region, the disturbance fields in the outer region may be expanded in the form: $\mathbf{u}=\mathbf{1_z}+Ri_v^{\frac{1}{3}}\mathbf{\tilde u}$, $p=p_{\infty}+Ri_v^{\frac{2}{3}}\tilde{p}$ and $\rho_f=\tilde{\rho_f}$, and one obtains the following equations for $\mathbf{\tilde u}$, $\tilde p$ and $\tilde \rho_f$:

\begin{equation} \label{oregion:largePe1}
\tilde \nabla. \mathbf{\tilde u}=0,
\end{equation}
\begin{equation} \label{oregion:largePe2}
-\alpha_\infty \frac{\partial \mathbf{\tilde u}}{\partial \tilde z}= -\tilde \nabla \tilde p +\tilde \nabla^2 \mathbf{\tilde u}-[\tilde {\rho_f}+6\pi\delta(\mathbf{ \tilde r})]\mathbf{1_z},
\end{equation}
\begin{equation} \label{oregion:largePe3}
-\mathbf{1_z}.\tilde u+ \frac{\partial\tilde{\rho_f}}{\partial \tilde z} = \beta_\infty \tilde \nabla^2 \tilde{\rho_f}.
\end{equation}
Here, $\alpha_\infty =\frac{Re}{Ri_v^{1/3}}$is the large-$Pe$ analog of $\alpha_0$, with $\beta_\infty^{-1}= \frac{Pe}{Ri_v^{1/3}}$ being the corresponding analog of $\beta_0$ above. In the Stokes stratification regime, corresponding to $Re \ll Ri_v^{\frac{1}{3}}$, one can set $\alpha_\infty$ in (\ref{oregion:largePe2}) to zero. Although our primary focus is on the limit $\beta_\infty \rightarrow 0\,(Pe\rightarrow \infty)$, retaining a small but finite $\beta_\infty$ turns out to be important for numerical convergence of the Fourier integrals below. As will also be seen below, the structure of the velocity and density fields, almost everywhere in the domain, is independent of $\beta_\infty$ provided the latter is small; in section \ref{jet:largePe}, however, it is shown that a small but finite $\beta_\infty$ crucially affects the structure of the fields right behind the translating sphere. Again, Fourier transforming, one obtains the velocity and density fields as the following integrals: 

\begin{equation}\label{eqn:axvel}
\mathbf{\tilde{u}} (\mathbf{\tilde{r}}) = \frac{-3}{4\pi^2} \int \frac{(ik_3+\beta_\infty k^2)k^2 (\mathbf{1_z}-\frac{k_3\mathbf{k}}{k^2})}{(ik_3+\beta_\infty k^2)k^4+k_t^2} e^{i\mathbf k.\mathbf{\tilde r}} d\mathbf{k},
\end{equation}
\begin{equation}\label{eqn:densdis}
{\tilde{\rho_f}}(\mathbf{\tilde{r}}) = \frac{-3}{4\pi^2} \int \frac{k_t^2}{(ik_3+\beta_\infty k^2)k^4+k_t^2} e^{i\mathbf k.\mathbf{\tilde r}} d\mathbf{k}.
\end{equation}

\section{Results and Discussion}\label{resul_disc}
Herein, we analyze the axial velocity and density disturbance fields, and the resulting streamline and iso-pycnal patterns in both the diffusion and convection dominant limits by using a combination of numerics (Gauss-Legendre quadrature integration) and far-field asymptotics. As already mentioned in the introduction, the results in both limits are for the case of buoyancy forces being dominant\,(the Stokes stratification regime), corresponding to $\alpha_0, \alpha_\infty \ll 1$. The role of weak inertial effects is discussed, via scaling arguments towards the end of this section.

\subsection{Diffusion-dominant limit ($Pe \ll 1$)}\label{diffdom_results}
\citet{list_laminar_1971} used residue theory to enable the reduction of the velocity and density fields to one-dimensional integrals for both the two and three-dimensional cases.  We use a different method where the disturbance fields are reduced to two-dimensional integrals; importantly, and unlike \citet{list_laminar_1971}, this method is applicable in both the diffusion and convection dominant limits. The Fourier integrals for the velocity and density disturbance fields, given by (\ref{eqn:axvellow}) and (\ref{eqn:densdislow}), are expressed in a spherical coordinate system with its polar axis aligned with the translation direction. The integral over the azimuthal angle ($\phi$) can be carried out analytically, and the resulting two dimensional integrals for the fields are given by:
\begin{equation}\label{eq:4v}
\bar{u}_{z}=\frac{-3}{2\pi}\int_0^\infty dk \int_0^\pi d\theta \frac{k^4\sin^3 \theta J_0(k\bar{r}_t\sin \theta)e^{ik\bar{z}\cos\theta}}{(k^4+\sin^2\theta)},
\end{equation}
\begin{equation}\label{eq:4}
\bar{\rho}_{f}=\frac{-3}{2\pi}\int_0^\infty dk \int_0^\pi d\theta \frac{k^2\sin^3 \theta J_0(k\bar{r}_t\sin \theta)e^{ik\bar{z}\cos\theta}}{(k^4+\sin^2\theta)},
\end{equation}
where $J_0(x)$ is the zeroth order Bessel function of the first kind. Note that since the problem is axisymmetric, the fields are written as functions of ($\bar r_t,\bar z$) with $\bar r_t$ and $\bar z$ being the distances along and orthogonal to the direction of translation. Not including the complex exponential, the Fourier integrand for the density disturbance field in (\ref{eq:4}) decays as $\frac{1}{k^{5/2}}$ for large $k$, while that for the axial velocity in (\ref{eq:4v}) only decays as $\frac{1}{k^{1/2}}$; the latter slow decay reflects the $1/r$-decay in physical space\,(for small $r$ corresponding to the inner region) of the Stokeslet. As a result, an accurate evaluation of (\ref{eq:4v}) relies essentially on cancellation induced by the complex Fourier exponential. In order to facilitate numerical evaluation, we therefore separate out the Stokeslet contribution, writing the axial velocity integral above as:

\begin{equation}\label{eq:5}
\mathbf{\bar{u}}_{z}=\frac{-3(2+\frac{\bar{r}_t^2}{\bar{z}^2})}{4\abs{\bar{z}}(1+\frac{\bar{r}_t^2}{\bar{z}^2})^{\frac{3}{2}}}+\frac{3}{2\pi}\int_0^\infty dk \int_0^\pi d\theta \frac{\sin^5\theta J_0(k\bar{r}_t\sin \theta)\cos(k\bar{z}\cos\theta)}{(k^4+\sin^2\theta)}
\end{equation}
%\begin{align}
%\bar{u}_{z_1}&=\frac{-3}{2\pi} \int_0^\infty dk \int_0^\pi d\theta \sin^3\theta J_0(k\bar{r}_t\sin \theta) e^{ik\bar{z}\cos\theta} \nonumber \\
%&+\frac{3}{2\pi}\int_0^\infty dk \int_0^\pi d\theta \frac{\sin^5\theta J_0(k\bar{r}_t\sin \theta)e^{ik\bar{z}\cos\theta}}{(k^4+\sin^2\theta)}
%\end{align}
where the Fourier integrand in $(\ref{eq:5})$ now decays as $\frac{1}{k^{9/2}}$ for large $k$, and we have replaced the complex exponential by the cosine on account of symmetry\,(an analogous replacement applies to (\ref{eq:4})). The Stokes streamfunction characterizing the axisymmetric flow field may be found from the axial velocity by using the relation $\bar{u}_z=\frac{1}{\bar{r}_t}\frac{\partial \bar{\psi}}{\partial\bar{r}_t}$ and is given by:

\begin{equation}\label{eq:6}
\mathbf{\bar{\psi}}_s=\frac{-3\bar{r}_t^2}{4(\bar{r}_t^2+\bar{z}^2)^{\frac{1}{2}}}+\frac{3\bar{r}_t}{2\pi}\int_0^\infty dk \int_0^\pi d\theta \frac{\sin^4\theta J_1(k\bar{r}_t\sin \theta)\cos(\bar{z}\cos\theta)}{k(k^4+\sin^2\theta)}
\end{equation}

\begin{figure}
	\centerline{\includegraphics[scale=1]{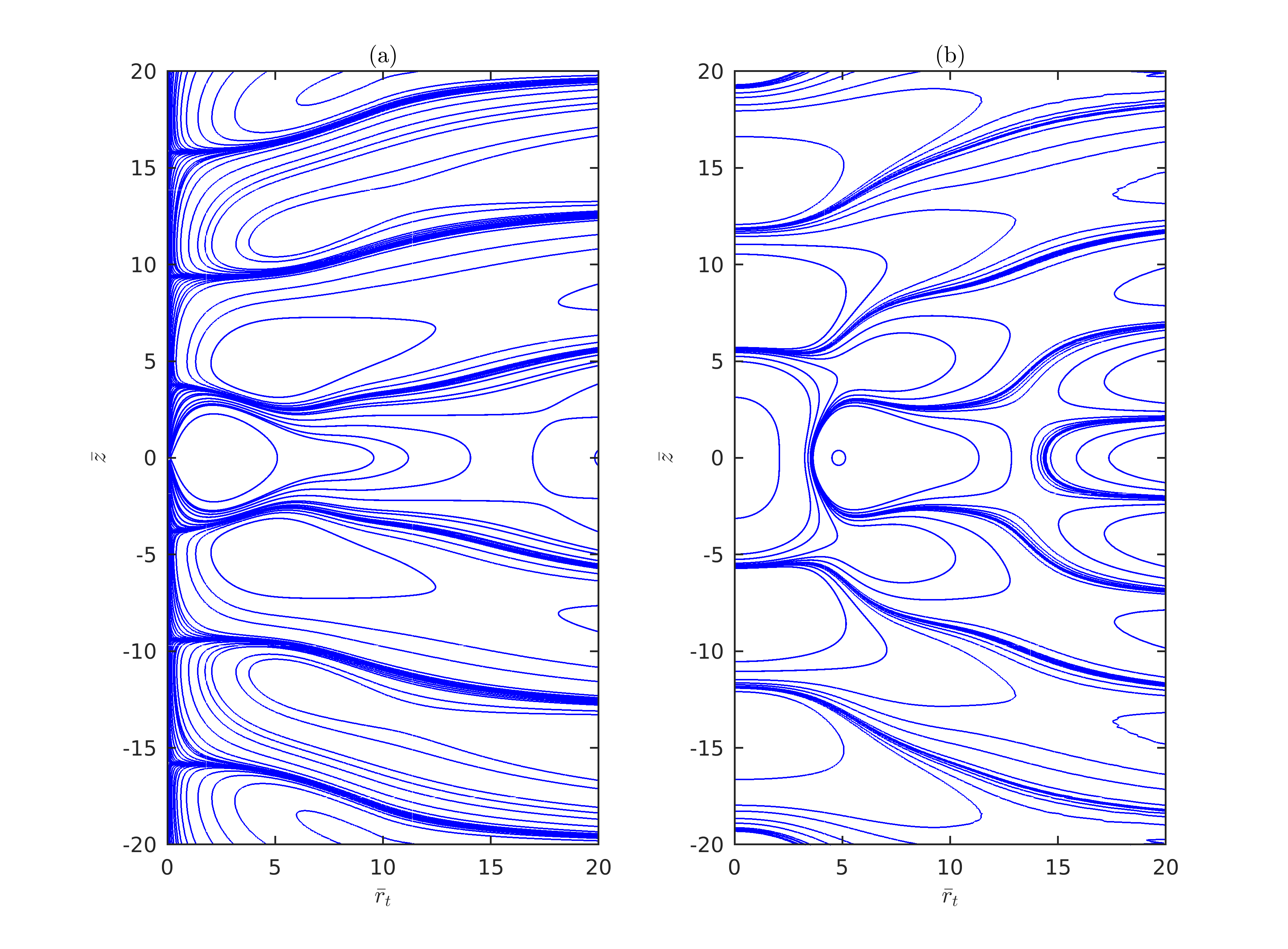}}% Images in 100% size
	\caption{(a) Streamlines and (b) Iso-pycnals for a translating sphere in a linearly stratified fluid in the diffusion dominant limit\,($Pe = 0$); in the point-particle approximation used, the sphere is at the origin and moving vertically downward.}
	\label{fig:lpcombined}
\end{figure}

The density disturbance and axial velocity fields, and the Stokes streamfunction, given by (\ref{eq:4}), (\ref{eq:5}) and (\ref{eq:6}), respectively, are evaluated using Gaussian quadrature. The instantaneous streamline pattern and iso-pycnals, in a reference frame with a far-field quiescent ambient, are shown in figure \ref{fig:lpcombined}. Both the disturbance velocity and density fields are seen to be fore-aft symmetric, as is evident from the cosine in (\ref{eq:5}) and (\ref{eq:6}). As originally found by \citet{list_laminar_1971} and \citet{ardekani_2010}, buoyancy forces suppress the long-ranged vertical motion associated with the Stokeslet at large distances, leading to the development of recirculating cells aligned with the direction of stratification, and wherein the motion is predominantly horizontal. Interestingly and perhaps surprisingly\,(if one's intuition is based on the cellular disturbance flow fields set up internal gravity waves in an unbounded stratified ambient), the far-field analysis in the next subsection shows the number of such cells to be finite, likely on account of the neglect of inertial/convection effects.

\subsubsection{Far-field analysis} \label{farfield:lowPe}
At large distances, as already mentioned, one expects the motion to be largely in the horizontal direction. As a consequence, one expects the characteristic length scale in the vertical direction to be much smaller than that along the horizontal - this is already evident from the rather small aspect ratios of the recirculating cells in figure \ref{fig:lpcombined}. Thus, the Fourier integrals in (\ref{eqn:axvellow}) and (\ref{eqn:densdislow}), for length scales large compared to $O(Ri_vPe)^{-1/4}$, may be simplified using $k_3 \gg k_t$, leading to:
\begin{equation}\label{fflpeq1}
\mathbf{\bar{u}} (\mathbf{\bar{r}}) = -\frac{3}{4\pi^2}\int \frac{k^4 (\mathbf{1_z}-\frac{k_3\mathbf{k}}{k^2})}{(k_3^6+k_t^2)} e^{i\mathbf k.\mathbf{\bar r}} d\mathbf{k},
\end{equation}
\begin{equation}\label{fflpeq2}
{\bar{\rho_f}} (\mathbf{\bar{r}}) = -\frac{3}{4\pi^2}\int \frac{k_t^2}{(k_3^6+k_t^2)} e^{i\mathbf k.\mathbf{\bar r}} d\mathbf{k}, 
\end{equation}
which may, via contour integration in the complex-$k_3$ plane, be reduced to one-dimensional integrals written in terms of the similarity variable $\eta = \frac{\bar z}{\bar {r_t}^{\frac{1}{3}}}$; see Appendix \ref{appendix1} for details. These integrals are only functions of $\abs {\eta}$, and are given by:

\begin{equation}\label{lpffweq1}
\bar{u}_{z}=\frac{-9i}{\bar{r}_t^3}\int_0^\infty p^8 J_0(p^3)\left(lq_1^2e^{iq_1p\abs{\eta}}+mq_2^2e^{iq_2p\abs{\eta}}+nq_3^2e^{iq_3p\abs{\eta}}\right)dp, 
\end{equation}
\begin{equation}\label{lpffweq3}
\bar{\rho}_{f}=\frac{-9i}{\bar{r}_t^{\frac{7}{3}}}\int_0^\infty p^6 J_0(p^3)\left(le^{iq_1p\abs{\eta}}+me^{iq_2p\abs{\eta}}+ne^{iq_3p\abs{\eta}}\right)dp,
\end{equation}
\begin{equation}\label{lpffweq2}
\bar{u}_{{r_t}}=\frac{-9\sign{(\eta)}}{\bar{r}_t^{\frac{7}{3}}}\int_0^\infty p^6 J_1(p^3)\left(lq_1^3e^{iq_1p\abs{\eta}}+mq_2^3e^{iq_2p\abs{\eta}}+nq_3^3e^{iq_3p\abs{\eta}}\right)dp.
\end{equation}

The above self-similar forms point to the existence of a thin axisymmetric wake bracketing the horizontal plane containing the settling sphere, in the far-field, whose vertical extent grows as $z \propto (Ri_vPe)^{-\frac{1}{6}}{r}_t^{\frac{1}{3}}$, where $z$ and $r_t$ are now in units of $a$; the disturbance fields are negligibly small outside the wake. Even within the wake, it can be seen from (\ref{lpffweq1}-\ref{lpffweq2}) that the disturbance fields exhibit a more rapid decay of the velocity field relative to the $O(1/r)$ decay of the Stokeslet, reinforcing the fact that buoyancy forces screen the originally long-ranged Stokesian fields. Nevertheless, the velocity and density fields in the diffusion-dominant limit are fore-aft symmetric as can be seen from the above expressions, and as evident from figure \ref{fig:lpcombined}. The one dimensional integrals in (\ref{lpffweq1}-\ref{lpffweq2}) are readily evaluated by using numerical integration, and furthermore, the large-$\eta$ asymptotes, obtained from using the small argument asymptote for the Bessel function in the integrands, are given by $\bar{u}_{z}\approx\frac{181440}{\bar{r}_t^3\abs{\eta}^9}$, $\bar{u}_{{r_t}}\approx\sign{(\eta)}\frac{816480}{\bar{r}_t^{7/3}\abs{\eta}^{10}}$, and $\bar{\rho}_{f}\approx\frac{-3240}{\bar{r}_t^{7/3}\abs{\eta}^{7}}$.

\begin{figure}
	\centerline{\includegraphics{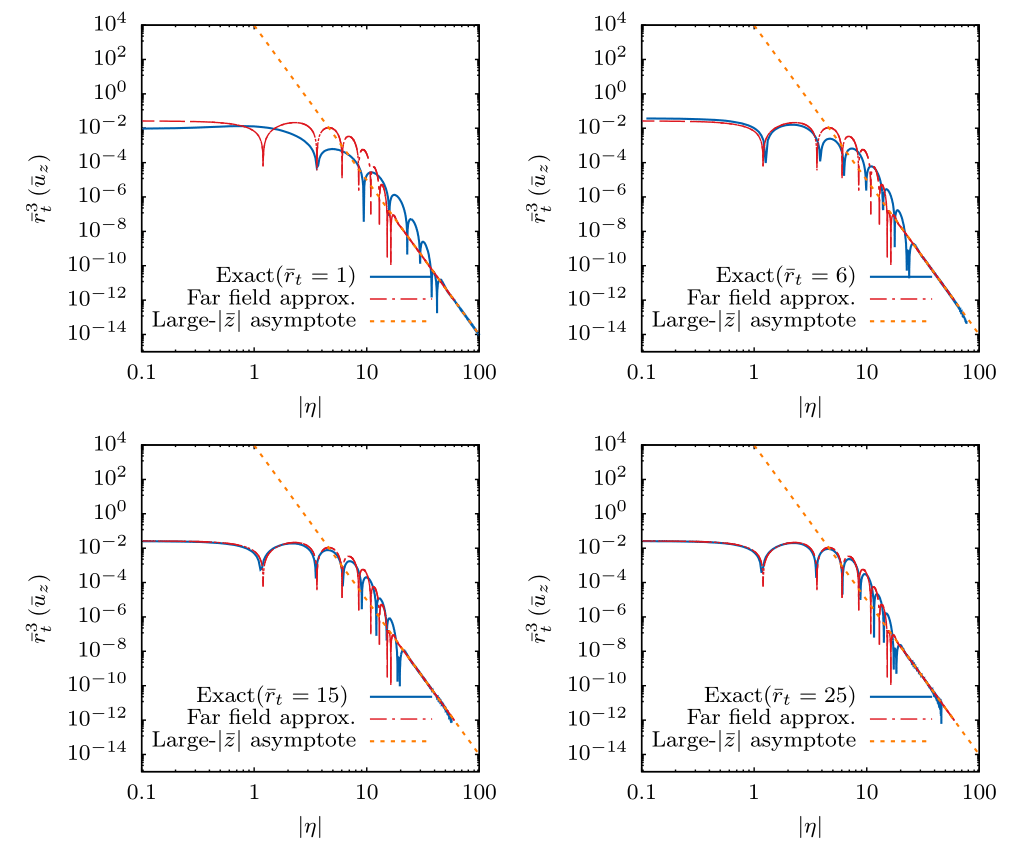}}% Images in 100% size
	\caption{The axial velocity profiles in the diffusion-dominant limit\,($Pe = 0$): comparison between the exact numerical profiles and the far-field approximation (given by (\ref{lpffweq1}) for various $\bar{r}_t$'s; in each of the plots, the large-$\eta$ analytical asymptote is shown as a dashed orange line.}
	\label{fig:lpfullfar}
\end{figure}

The comparison between the one-dimensional profiles of the axial velocity field, obtained from the exact calculations above (that led to the streamline pattern in figure \ref{fig:lpcombined}), and those obtained from the far-field self-similar approximation given by (\ref{lpffweq1}) are shown in figure \ref{fig:lpfullfar} for various $\bar{r}_t$'s. Based on the self-similar form given by (\ref{lpffweq1}), the figures plot $\bar{r}_t^3 \bar{u}_z$ as a function of $|\eta|$, as a result of which the far-field approximation shown in the figures remains invariant to a change in $\bar{r}_t$. In the log-log plots shown, the zero-crossings of the axial velocity (which roughly correlate to the boundaries between recirculating cells) appears as sharp dips\,(to negative infinity). While there exist significant differences between the numerical and far-field predictions for $\bar{r}_t$'s of order unity, the agreement improves with increasing $\bar{r}_t$, and there is near-quantitative agreement for the largest $\bar{r}_t\,(=25)$. Importantly, the number of zero crossings\,(eight) in the exact field appears independent of $\bar{r}_t$, and is the same as that in the far-field approximation; note that the streamline pattern in figure \ref{fig:lpcombined} includes only three of the eight zero crossings for $\bar{r}_t = 25$. The finite number of zero crossings seen in figure \ref{fig:lpfullfar}, as mentioned above, points to a finite number of recirculating cells in the outer region. Finally, for $\bar{r}_t$'s greater than that corresponding to the final zero crossing, the axial velocity profiles conform to the algebraic asymptote given above viz.\! $\bar{r}_t^3\bar{u}_{z}\approx\frac{181440}{\abs{\eta}^9}$, and shown as the dashed orange line in figure \ref{fig:lpcombined}. A scenario analogous to that described above prevails for the density disturbance field.

\begin{figure}
	\centerline{\includegraphics{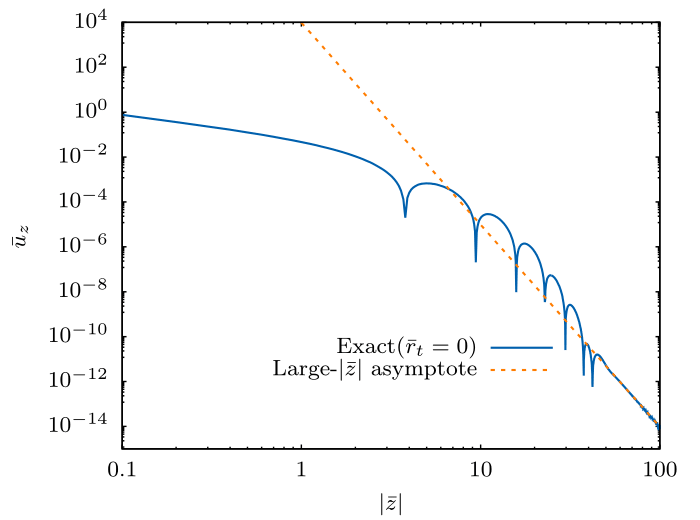}}% Images in 100% size
	\caption{Axial velocity, as a function of $\bar{z}$, for $\bar{r}_t = 0$\,(the translation axis), in the diffusion dominant limit\,($Pe = 0$); the large-$\bar{z}$ asymptote is shown as a dashed orange line.}
	\label{fig:lpfullasymrt0}
\end{figure}

As one approaches the translation axis, that is, for $\bar{r}_t \rightarrow 0$, $\eta$ becomes asymptotically large for any finite $\bar{z}$, and only the large-$\eta$ asymptotes are of relevance. On substituting for $\eta$, the large-$\eta$ asymptotes for the axial velocity and density fields above are seen to be independent of $\bar{r}_t$, being functions of only $|\bar{z}|$, suggesting that these asymptotes remain valid far-field\,(large $|\bar{z}|$) approximations even along the translation axis\,(the stagnation streamline). The radial velocity is, of course, zero along the stagnation streamline, with the large-$\eta$ approximation given above being $O(\bar{r}_t)$ for small $\bar{r}_t$. In figure \ref{fig:lpfullasymrt0}, we compare the exact axial velocity field for $\bar{r}_t = 0$, again obtained numerically, with the large-$\eta$ asymptote that is now proportional to $\bar{z}^{-9}$. Although the locations of the (seven)\,zero-crossings of the exact profile can no longer be predicted, the far-field algebraic decay nevertheless conforms to the asymptote above. It is worth noting that, the large-$z$ asymptote may also be obtained by directly setting $\bar{r}_t = 0$ in the exact expression for the axial velocity, giving:
\begin{align}
u_z&=\frac{-1}{4\pi z}+\frac{1}{2\pi^2}\int_0^{\pi/2}d\theta\int_0^\infty dk\;\frac{\sin^5\theta \cos(kz\cos \theta)}{k^4+\sin^2\theta}, \nonumber
\end{align}
which in turn may be reduced to the following one dimensional integral using residue integration:
\begin{equation}\label{eq:StagnationVel}
u_z =\frac{-1}{4\pi z}+\frac{1}{4\pi}\int_0^{\pi/2} d\theta \;e^{-z\cos\theta\sqrt{\frac{\sin\theta}{2}}}\cos(z\cos\theta\sqrt{\frac{\sin\theta}{2}}-\frac{\pi}{4})\sin^{7/2}\theta,
\end{equation}
a reduction only possible for $\bar{r}_t=0$. For large $\bar z$, the dominant contributions to the above integral arise from the neighborhood of the zeroes of $\cos\theta\sqrt{\frac{\sin\theta}{2}}$ - that is, $\theta=0$ and $\theta=\frac{\pi}{2}$. The contribution from $\theta=\pi/2$ exactly cancels the Stokeslet contribution\,(the first term in (\ref{eq:StagnationVel})). The second order contribution from $\theta=\pi/2$, and the leading order contribution from $\theta=0$, together, lead to the large-$\bar{z}$ asymptote above, which was originally given in \citet{fouxon_2014}.

\subsection{Convection dominant limit ($Pe \gg 1$)}\label{convecdom_results}
\begin{figure}
	\centerline{\includegraphics{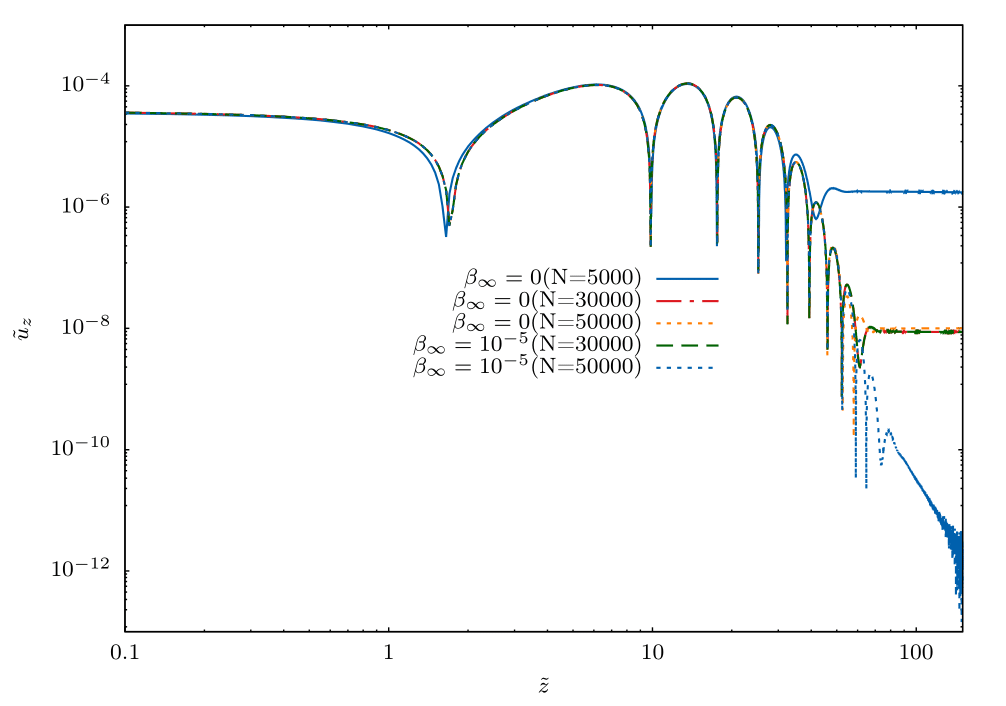}}% Images in 100% size %[width=0.8\linewidth,height=0.45\textheight]
	\caption{The comparison given highlights the importance of weak diffusive effects\,(small but finite $\beta_\infty$) in obtaining an accurate representation of the disturbance fields in the convection-dominant limit. The profiles for $\beta_\infty = 0$ asymptote to a spurious plateau regardless of $N$; here, $N$ represents the number of quadrature points used for numerical integration.}
	\label{betaConv_comparison}
\end{figure}

The Fourier integrals in the convection dominant limit are given by (\ref{eqn:axvel}) and (\ref{eqn:densdis}), and their simplification is analogous to the diffusion dominant case above. In a spherical coordinate system aligned with the translation direction, and after integration over the azimuthal angle, the residual two-dimensional integrals for the disturbance fields are given by:
\begin{equation}\label{eqn:axvel2}
\tilde{u}_{z}=\frac{-3(2+\frac{\tilde{r}_t^2}{\tilde{z}^2})}{4\abs{\tilde{z}}(1+\frac{\tilde{r}_t^2}{\tilde{z}^2})^{\frac{3}{2}}}+\frac{3}{2\pi}\int_0^\infty dk \int_0^\pi d\theta \frac{\sin^5\theta J_0(k\tilde{r}_t\sin \theta)e^{ik\tilde{z}\cos\theta}}{(ik^3\cos\theta+\beta_{\infty}k^4+\sin^2\theta)},
\end{equation}
\begin{equation} \label{eqn:densdis2}
\tilde{\rho_f}=\frac{-3}{2\pi}\int_0^\infty dk \int_0^\pi d\theta \frac{k^2\sin^3\theta J_0(k\tilde{r}_t\sin \theta)e^{ik\tilde{z}\cos\theta}}{(ik^3\cos\theta+\beta_{\infty}k^4+\sin^2\theta)},
\end{equation}
where, as for the diffusion-dominant case, we have separated out the Stokeslet contribution in (\ref{eqn:axvel2}) in the interests of numerical convergence. The Stokes streamfunction can be derived from the axial velocity as before and is given by 
\begin{equation} \label{Stokesstream:2}
\tilde{\psi}_s=\frac{-3\tilde{r}_t^2}{4(\tilde{r}_t^2+\tilde{z}^2)^{\frac{1}{2}}}+\frac{3\tilde{r}_t}{2\pi}\int_0^\infty dk \int_0^\pi d\theta \frac{\sin^4\theta J_1(k\tilde{r}_t\sin \theta)e^{ik\tilde{z}\cos\theta}}{k(ik^3\cos\theta+\beta_{\infty}k^4+\sin^2\theta)}
\end{equation}
Note from (\ref{eqn:axvel2}) and (\ref{eqn:densdis2}) that, although our interest is in the limit $\beta_\infty = 0$, corresponding to convection effects being infinitely dominant, we have nevertheless retained the terms proportional to $\beta_\infty$ in the Fourier integrands. This is because, on one hand, numerical convergence in the convection-dominant limit is considerably more difficult; a small but finite $\beta_\infty$ aids convergence of the quadrature integration especially at large distances from the sphere, and over most of the domain, as is evident from figure \ref{betaConv_comparison} where we compare the numerically evaluated axial velocity  profiles for $\beta_\infty = 0$ and $10^{-5}$ for varying number of quadrature points. The detailed explanation of the nature of this profile appears later, but it may nevertheless be seen that the $\beta_\infty = 0$ profile deviates from the true profile, asymptoting to a spurious plateau beyond a certain $\tilde{z}$. There is only a modest effect of quadrature resolution on this threshold $\tilde{z}$, and as a result, for $\beta_\infty = 0$, the eventual algebraic decay regime remains numerically inaccessible regardless of the number of quadrature points. On the other hand, and more importantly, the structure of both the velocity and density fields behind the translating sphere, in the vicinity of the rear stagnation streamline, depends crucially on $\beta_\infty$ being non-zero; the density field in particular is logarithmically singular on the rear stagnation streamline for $\beta_\infty = 0$.\\

\begin{figure}
	\centerline{\includegraphics[scale=1]{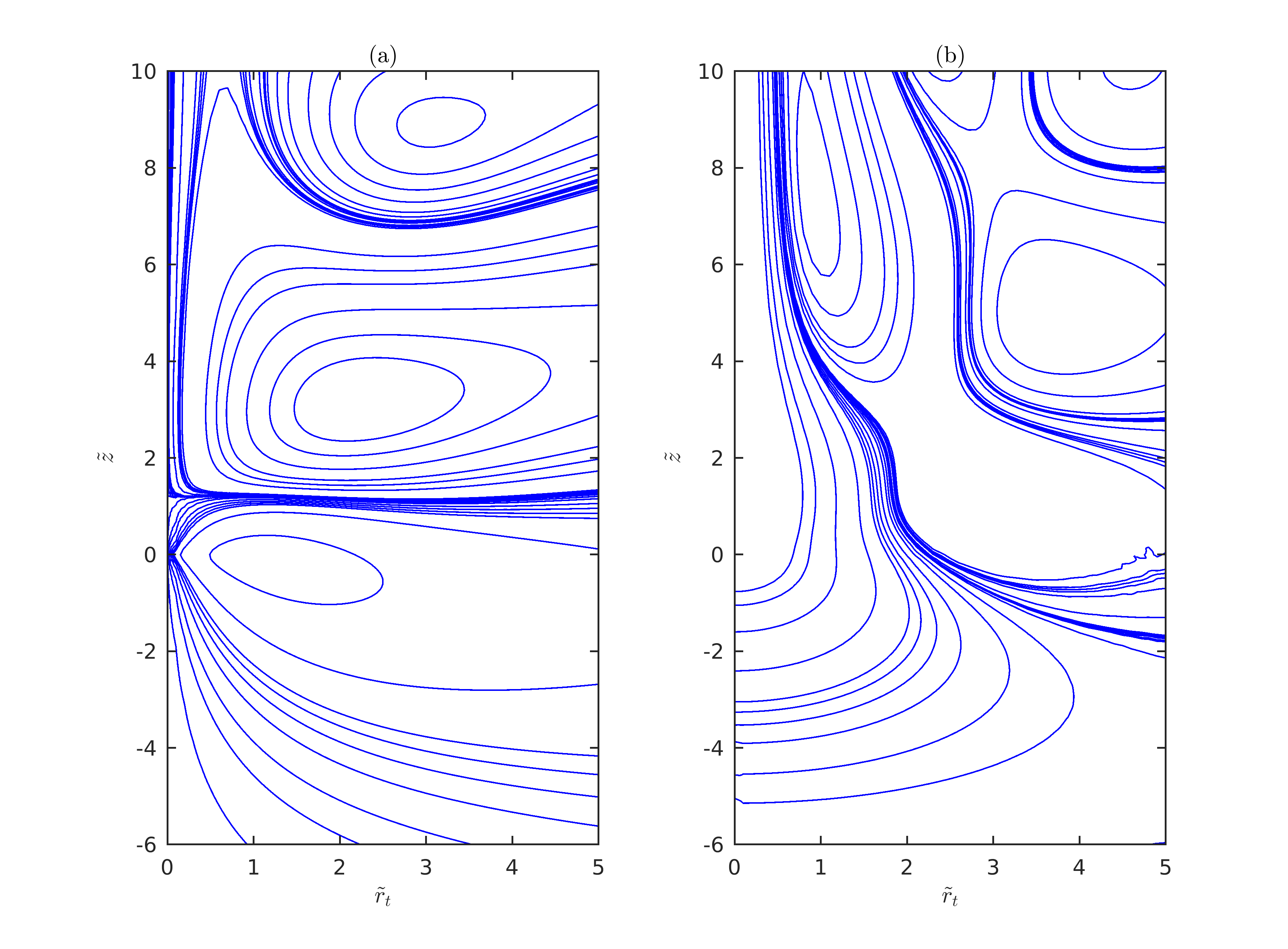}}% Images in 100% size %[width=1.15\linewidth,height=0.45\textheight]
	\caption{(a) Streamlines and (b) Iso-pycnals for a translating sphere in a linearly stratified fluid, in the convection dominant limit\,($\beta_\infty = 10^{-5}$), in the Stokes stratification regime\,($Re \ll Ri_v^{\frac{1}{3}}$); in the point-particle approximation used, the sphere is at the origin and moving vertically downward.}
	\label{fig:hpstream_beta105}
\end{figure}

Figure \ref{fig:hpstream_beta105} shows the streamline pattern and the isopycnal contours for the smallest $\beta_\infty\,(=10^{-5})$ accessed in our calculations. The limited spatial extent here, in comparison to figure \ref{fig:lpcombined}, is on account of the numerical difficulties involved in calculating the farfield isopycnals; the streamline pattern alone, over a larger spatial extent, appears below in figure \ref{fig:hpstream}a. Nevertheless, the profound asymmetry of both the streamline and iso-pycnals patterns is readily evident. This asymmetry may be anticipated from the integral expressions in (\ref{eqn:axvel2}-\ref{Stokesstream:2}) where, unlike the $Pe = 0\,(\beta_\infty = \infty)$ limit, one may no longer replace the complex exponential by a Cosine. Apart from the different shapes and numbers of the recirculating cells in front of and behind the translating sphere, evident from the figure, there is also the appearance of a radially localized but vertically extended structure, in the streamline pattern, in the rear. As will be seen, this corresponds to a buoyant reverse jet that develops behind the particle with decreasing $\beta_\infty$. The far-field analysis below points to both a stratification-induced wake in the convection-dominant limit with a structure that is insensitive to $\beta_\infty$\,(for $\beta_\infty \ll 1$); and the buoyant reverse jet mentioned above whose structural features depend essentially on $\beta_\infty$; these are analyzed in separate subsections.

\subsubsection{Far-field wake analysis} \label{farfieldwake:largePe}

\begin{figure}
	\centerline{\includegraphics{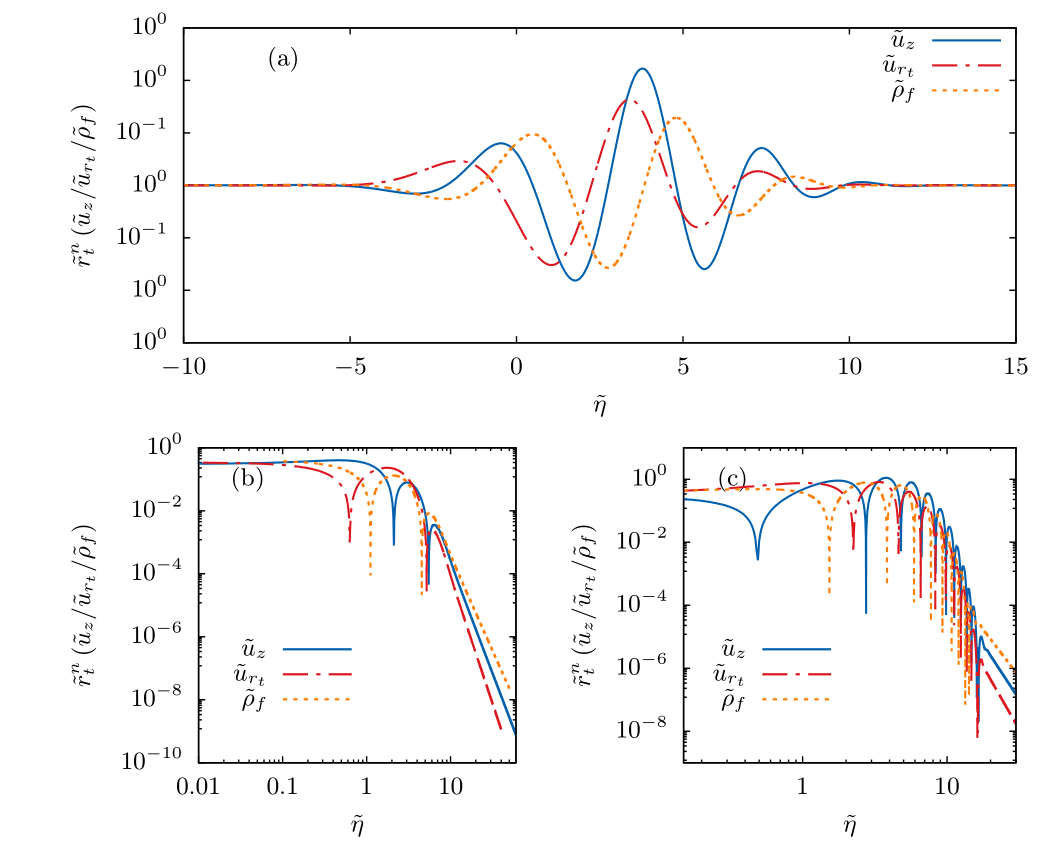}}% Images in 100% size
	\caption{The axial velocity, the radial velocity and density disturbance profiles, within the far-field wake region, in the convection dominant limit\,($Pe \gg 1$) pertaining to the Stokes stratification regime: $(a)$ the disturbance fields on a linear scale; the absolute value of the disturbance fields on a logarithmic scale for (b) negative $\tilde \eta$ and $(c)$ for positive $\tilde \eta$; here, $n = 14/5$, $11/5$ and $12/5$ for $u_z$, $u_{r_t}$ and $\rho_f$, respectively. The aforementioned wake includes the plane of the settling sphere, and grows in vertical extent as $z \propto Ri_v^{\frac{2}{15}}{r}_t^{\frac{2}{5}}$.}
	\label{fig:hpffwsim}
\end{figure}

Similar to the diffusion-dominant case analyzed in section \ref{farfield:lowPe}, the expected dominance of horizontal motion for distances large compared to $Ri_v^{-\frac{1}{3}}$ points to the assumption $k_3 \gg k_t$ being applicable to the Fourier integrals in (\ref{eqn:axvel}) and (\ref{eqn:densdis}), when characterizing fluid motion in a far-field wake region. The original Fourier integrals, in this limit,reduce to:
\begin{equation} \label{highpevelffw}
\mathbf{\tilde{u}} (\mathbf{\tilde{r}}) = \frac{-3}{4\pi^2} \int \frac{ik_3k_t^2}{(ik_3^5+k_t^2)} e^{i\mathbf k.\mathbf{\tilde r}} d\mathbf{k},
\end{equation}

\begin{equation} \label{highpedensffw}
{\tilde{\rho_f}}(\mathbf{\tilde{r}}) = \frac{-3}{4\pi^2} \int \frac{k_t^2}{ik_3^5+k_t^2} e^{i\mathbf k.\mathbf{\tilde r}} d\mathbf{k},
\end{equation}
where we have set $\beta_\infty = 0$ which, as will be seen, is justified everywhere in the domain except in the vicinity of the rear stagnation streamline. The integrals in (\ref{highpevelffw}) and (\ref{highpedensffw}) may be reduced to the following one-dimensional integrals, written in terms of the similarity variable $\eta = \frac{\tilde z}{\tilde {r_t}^{\frac{2}{5}}}$, via contour integration\,(see Appendix \ref{appendix2} for details):
\begin{align}\label{hpffweq2}
\tilde {u}_{z}&=-\frac{15i}{2\tilde{r_t}^{14/5}}\int_0^\infty m^6J_0[m^{5/2}][Q_1q_1e^{iq_1m\eta}+Q_2q_2e^{iq_2m\eta}+Q_3q_3e^{iq_3m\eta}]dm \textrm{ } (\textrm{for } \tilde \eta>0), \nonumber \\
&=\frac{15i}{2\tilde{r_t}^{14/5}}\int_0^\infty m^6J_0[m^{5/2}][Q_4q_4e^{iq_4m\eta}+Q_5q_5e^{iq_5m\eta}]dm \textrm{ } (\textrm{for } \tilde \eta<0),
\end{align}
\begin{align}\label{hpffweq3}
\tilde {\rho_f} &=-\frac{15}{2\bar{r_t}^{12/5}}\int_0^\infty m^5J_0[m^{5/2}][Q_1e^{iq_1m\eta}+Q_2e^{iq_2m\eta}+Q_3e^{iq_3m\eta}]dm \textrm{ } (\textrm{for } \tilde \eta>0), \nonumber\\
&=\frac{15}{2\tilde{r_t}^{12/5}}\int_0^\infty m^5J_0[m^{5/2}][Q_4e^{iq_4m\eta}+Q_5e^{iq_5m\eta}]dm \textrm{ } (\textrm{for } \tilde \eta<0),
\end{align}
\begin{align}\label{hpffweq1}
\tilde u_{r_t}&=-\frac{15}{2\tilde {r_t}^{11/5}}\int_0^\infty m^{9/2}J_1[m^{5/2}][Q_1q_1^2e^{iq_1m\eta}+Q_2q_2^2e^{iq_2m\eta}+Q_3q_3^2e^{iq_3m\eta}]dm \textrm{ } (\textrm{for } \tilde \eta>0), \nonumber \\
&=\frac{15}{2\tilde {r_t}^{11/5}}\int_0^\infty m^{9/2}J_1[m^{5/2}][Q_4q_4^2e^{iq_4m\eta}+Q_5q_5^2e^{iq_5m\eta}]dm \textrm{ } (\textrm{for } \tilde \eta<0).
\end{align}
Here, the $Q_n$'s and $q_n$'s ($n=1,2,3,4,5$) are complex-valued constants given in Appendix \ref{appendix2}. The fore-aft asymmetry implies that one has different asymptotic approximations depending on the sign of $\tilde{\eta}$\,(or $\tilde{z}$). Nevertheless, the above self-similar forms point to a far-field wake, that includes the horizontal plane containing the settling sphere, and whose vertical extent grows as $z \propto Ri_v^{\frac{2}{15}}{r}_t^{\frac{2}{5}}$, with $z$ and $r_t$ being measured in units of $a$. The axial and radial velocity profiles, and the density disturbance profiles, obtained from a numerical evaluation of the one-dimensional integrals above, are shown both on the linear and logarithmic scales in figure \ref{fig:hpffwsim}. The logarithmic plot shows that while there are still only a finite number of zero crossings, similar to $Pe = 0$, they differ in number for negative and positive $\tilde{\eta}$, with fewer zero crossings for negative $\tilde{\eta}$. This implies fewer recirculating cells below the settling sphere, and is consistent with the streamline pattern in figure \ref{fig:hpstream_beta105}. Similar to the diffusion-dominant limit, one may obtain the large-$\eta$ asymptotic forms from (\ref{hpffweq2}-\ref{hpffweq1}) which govern the eventual algebraic decay of the disturbance fields beyond the final zero crossing; these are given by  $[\frac{3240}{r_t^{\frac{14}{5}}\tilde{\eta}^7},-\frac{2160}{r_t^{\frac{14}{5}}\tilde{\eta}^ 7}]$ for the axial velocity, $[\frac{11340}{r_t^{\frac{11}{5}}\tilde{\eta}^ 8},-\frac{7560}{r_t^{\frac{11}{5}}\tilde{\eta}^ 8}]$ for the radial velocity, and $[-\frac{540}{r_t^{\frac{12}{5}}\tilde{\eta}^ 6},\frac{360}{r_t^{\frac{12}{5}}\tilde{\eta}^ 6}]$ for the density disturbance, with the first and second members of each ordered pair corresponding to positive and negative $\tilde{\eta}$, respectively. These asymptotes, and the above approximate profiles based on the one-dimensional integrals above will be compared to the exact numerically evaluated disturbance fields below. 

The structure of the far-field wake may also be characterized in terms of the $\tilde{\eta}$-moments of the disturbance fields above. The motion being largely horizontal, it is the moments of the radial velocity field that are the most important. A calculation using the far-field approximation above\,(equation (\ref{hpffweq1})) shows that the zeroth and first moments of the radial velocity field in the wake vanish, and that the second moment, defined as $\int_{-\infty}^\infty \tilde{\eta}^2 \tilde{u}_{r_t} d\tilde{\eta} = 6$, is the first non-trivial member of the moment hierarchy\,(interestingly, this may also be seen from direct neglect of the viscous term $ik_3k^4$ in the original Fourier integral (\ref{eqn:axvel}), and additionally setting $\beta_\infty = 0$; the radial velocity may now be obtained in terms of generalized functions as $\tilde u_{r_t} =  \frac{3}{\tilde{r}_t}\delta''(z)$, that yields the same value for the second moment.). The moment-based characterization above offers an interesting contrast to the known solution for the motion induced by a sphere settling through a linearly stratified ambient in the linearized inviscid approximation, when stratification forces are (infinitely)\,dominant. As shown in (\cite{vladimirov1991}), the motion is strictly horizontal and restricted to an infinite horizontal slab whose upper and lower planes bound the sphere. Within this slab, the fluid moves radially inward\,(outward) in the rear\,(front) half of the translating sphere. The nature of this motion is easily understood from the changing size of the sphere cross-section in a given horizontal plane, and the requirement of satisfying the impenetrability condition at the sphere surface. In two dimensions\,(that is, a settling cylinder), the horizontal velocity field is a constant, while in three dimensions\,(a settling sphere), the motion would have a $1/r_t$-dependence consistent with incompressibility. Such a motion corresponds has a dipolar character with a non-trivial first moment for the radial velocity. In contrast, as already seen, the structure of the far-field wake above does not exhibit the aforementioned structure. This is because although the Stokeslet in the inner reigon has a radial component consistent with the symmetry of the linearized inviscid solution above\,(directed inward behind the sphere and outward in front of it), the force associated with the Stokeslet is screened by the buoyancy forces induced by the density perturbation, in a surrounding volume with a dimension of $O(Ri_v^{-\frac{1}{3}})$. As a result, the wake velocity field on length scales much larger than $O(Ri_v^{-\frac{1}{3}})$, has the symmetry pertaining to a force-dipole consisting of the original Stokeslet and an effective upward force arising from the aforementioned volumetric distribution of induced buoyancy forces.

\subsubsection{Far-field jet analysis}  \label{jet:largePe}
As for the diffusion-dominant case, the large-$\eta$ asymptotes for the axial velocity and density disturbance fields in the convection-dominant limit, given above, are seen to be independent of $\bar{r}_t$, with the radial disturbance field being $O(\bar{r}_t)$ for $\tilde{z} \rightarrow 0$. Thus, one expects the large-$\eta$ asymptotes to continue to remain valid at sufficiently large distances\,(large $\tilde{z}$) along the stagnation streamline\,($\tilde{z} = 0$). This remains true for the front stagnation streamline, with $\tilde{u}_z = -\frac{2160}{\tilde{z}^7}$ and $\rho_f = \frac{360}{\tilde{z}^6}$ for large negative $\tilde{z}$. Although we don't go into any detail, these far-field asymptotes may also be derived directly from the exact expressions via residue integration, as seen in (\ref{eq:StagnationVel}) for $Pe = 0$.

The wake approximation in the earlier subsection, and therefore, the large-$\eta$ approximations derived from it, are no longer valid in the vicinity of the rear stagnation streamline. The pronounced asymmetry in the streamline pattern in figure \ref{fig:lpcombined}, and the predominantly vertical motion behind the sphere, are already indicative of the breakdown of the wake approximation.The neighborhood of the rear stagnation streamline, at large distances, corresponds to large positive $\tilde{z}$ and small $\tilde{r}_t$, which in Fourier space is equivalent to $k_3 \ll k_t$ - the opposite of the wake-approximation developed above. This reduces the original Fourier integrals to the following approximate forms:
\begin{equation} \label{jetintegral:1}
\mathbf{\tilde{u}} (\mathbf{\tilde{r}}) = -\frac{6\pi i}{8\pi^3}\int \frac{k_3k_t^2 (\mathbf{1_z}-\frac{k_3\mathbf{k}}{k^2})}{(ik_3k_t^4+k_t^2+\beta_\infty k_t^6)} e^{i\mathbf k.\mathbf{\tilde r}}d\mathbf{k},
\end{equation}
\begin{equation} \label{jetintegral:2}
{\tilde{\rho_f}} (\mathbf{\tilde{r}}) = -\frac{6\pi}{8\pi^3}\int \frac{k_t^2}{(ik_3k_t^4+k_t^2+\beta_\infty k_t^6)}e^{i\mathbf k.\mathbf{\tilde r}}d\mathbf{k},
\end{equation}
where, unlike the wake-approximation above, we retain the $O(\beta_\infty)$ terms in the integrands, in anticipation of the fact that the reverse jet we find below has a structure that crucially depends on $\beta_\infty$ even in the limit $\beta_\infty \ll 1$. The integrals in (\ref{jetintegral:1}-\ref{jetintegral:2}) can be further simplified by contour integration in the complex-$k_3$ plane. From the denominator of the integrand in (\ref{jetintegral:1}-\ref{jetintegral:2}) one notes that the only pole exists in the upper half of the complex plane, being given by $k_3=i\frac{\beta_\infty k_t^4+1}{k_t^2}$. This pole contributes only for positive $z$, when one closes the contour via a semi-circle\,(of an infinite radius) in the upper half of the plane. Performing the integral over the azimuthal angle, and accounting for the contribution of the aforementioned pole in the $k_3$-integration, the axial velocity and density disturbance fields can be reduced to the following one-dimensional integrals:
\begin{align}
\tilde{u}_{z}=3\int_0^\infty \frac{J_0(k_t\tilde{r}_t)e^{-\tilde{z}(\beta_\infty k_t^2+\frac{1}{k_t^2})}}{k_t^3} dk_t, \label{jet:1Dintegral1} \\
\tilde{\rho}_{f}=-3\int_0^\infty \frac{J_0(k_t\tilde{r}_t)e^{-\tilde{z}(\beta_\infty k_t^2+\frac{1}{k_t^2})}}{k_t} dk_t,\label{jet:1Dintegral2}
\end{align}
For $\tilde{r}_t=0$, the integrals in (\ref{jet:1Dintegral1}) and (\ref{jet:1Dintegral1}) may be evaluated analytically, giving: 
\begin{align}
\tilde{u}_{z}=3\sqrt{\beta_\infty} K_1[2\sqrt{\beta_\infty}\tilde{z}], \label{eqn:axialfarjet1} \\
\tilde{\rho}_{f}=-3K_0[2\sqrt{\beta_\infty}\tilde{z}]. \label{eqn:axialfarjet2}
\end{align}
Here, $K_0$ and $K_1$ are zeroth and first order modified Bessel functions of the second kind, respectively. The crucial role of weak diffusion on the jet structure, as characterized by (\ref{eqn:axialfarjet1}) and (\ref{eqn:axialfarjet2}), may now be seen. Rather remarkably, on using the small-argument asymptote $K_1(z) \approx 1/z$ in  the limit $\beta_\infty \rightarrow 0$, (\ref{eqn:axialfarjet1}) is found to be independent of $\beta_\infty$ at leading order, reducing to $\tilde u_z \approx \frac{3}{2\tilde z}$. This implies that the axial velocity, although pointing in the reverse direction\,(that is, directed opposite to the translating sphere), still decays as $O(1/z)$, analogous to a Stokeslet, on length scales much larger than $O(Ri_v^{-\frac{1}{3}})$! In contrast, on using the small argument form $K_0 \approx - \ln z$, the density disturbance given by (\ref{eqn:axialfarjet2}) is seen to be logarithmically singular for $\beta_\infty \rightarrow 0$ for any positive $\tilde{z}$, pointing to a logarithmic singularity all along the rear stagnation streamline for $Pe = \infty$. The far-field behavior in this jet region changes fundamentally for any small but finite $\beta_\infty$. Now, there exists a second screening length across which the buoyant jet transitions from the $\frac{1}{z}$ decay above to a much faster exponential one, this arising from the exponentially decaying forms of the large-argument asymptotes of the modified Bessel functions above; likewise, the density disturbance transitions from the logarithmic form above, again to a far-field exponential decay. From (\ref{eqn:axialfarjet1}) and (\ref{eqn:axialfarjet2}), this second screening length is seen to be $O(\beta_\infty^{-\frac{1}{2}})$, in units of $Ri_v^{-\frac{1}{3}}$, or $O(Ri_v^{-\frac{1}{6}} Pe^{\frac{1}{2}})$ in units of $a$. The radial extent of the jet region may be seen from the earlier expressions (\ref{jetintegral:1}) and (\ref{jetintegral:2}). Setting $\beta_\infty  =0$, one notes that $ \tilde z \sim O(k_t^2) \sim O(\tilde r_t^{-2})$ for the argument of the exponential integrand to be of order unity. Thus, the reverse-Stokeslet behavior above is valid in a region with a radial extent $\tilde{r}_t \propto \tilde{z}^{-\frac{1}{2}}$ for $\beta_\infty=0$, suggesting that the buoyant jet narrows as $O(\tilde{z}^{-\frac{1}{2}})$, with increasing downstream distance, until the effects of diffusion become important. As shown above, the diffusive smearing of the jet, and the transition to an exponentially decaying reverse flow, occurs across a second screening length of $O(\beta_\infty^{-\frac{1}{2}})$ when the jet has a width of $O(\beta_\infty^{\frac{1}{4}})$, both in units of $Ri_v^{-\frac{1}{3}}$. Although, the existence of a rearward jet is well known for moderate Reynolds numbers, from earlier computations\,(see \citep{hanazaki_2009}), its appearance has been primarily attributed to the inertial effects\,(for instance, see \citep{eames_inviscid_1997}). The existence of such a jet, as predicted above in the Stokes stratification regime, therefore comes as a surprise. It is also worth emphasizing that, unlike the usual case of the laminar\,(or turbulent) wake or jet, the buoyant jet above conserves neither momentum nor mass flux; the absence of a net mass flux implies that the existence of a jet region doesn't affect drift volume estimates\,(see section \ref{Drift_estimates}).

\begin{figure}
	\centerline{\includegraphics[scale=1.05]{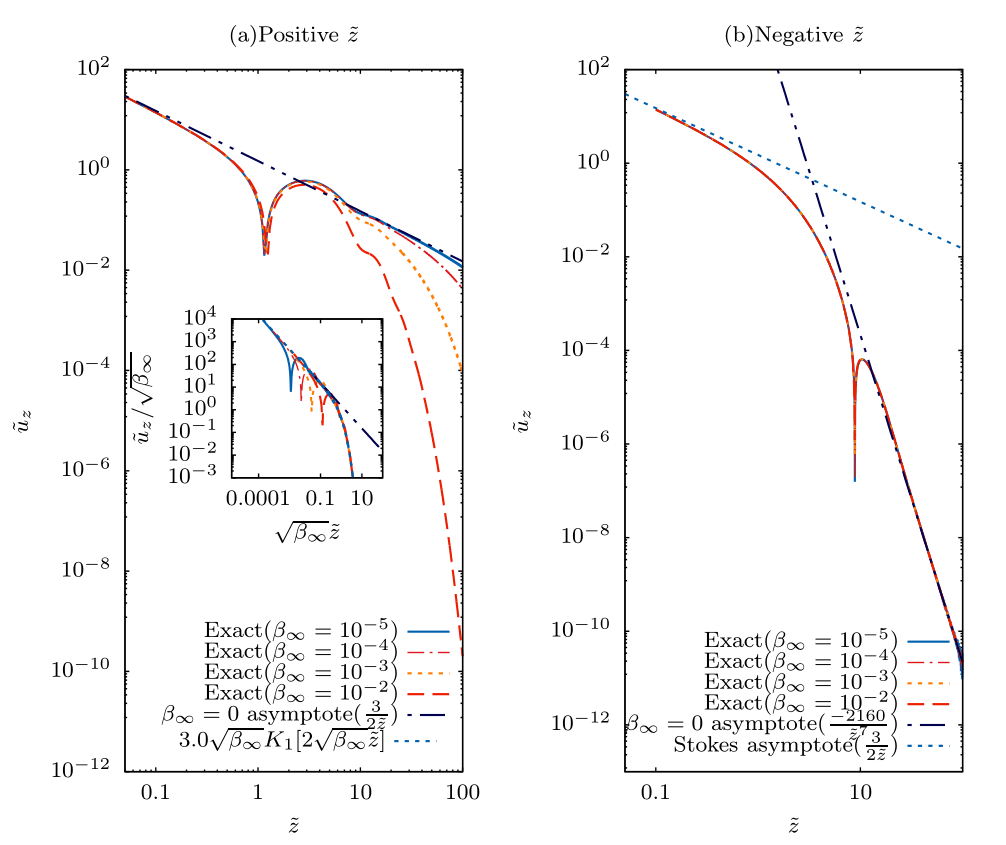}}% Images in 100% size
	\caption{The axial velocity field plotted along the stagnation streamline for both positive\,(the LHS plot) and negative $\tilde{z}$\,(the RHS plot), and for different small $\beta_\infty$. In the LHS plot, both the Stokeslet and reverse-Stokeslet asymptotes appear as the black dot-dashed line; the inset shows the collapse of the far-field profiles onto a common curve, when plotted as a function of $\beta_\infty^{\frac{1}{2}}\tilde{z}$, consistent with (\ref{eqn:axialfarjet1}). The RHS plot shows the transition from the near-field Stokeslet decay\,(blue dash-dotted line) to the far-field decay given by $-\frac{2160}{\tilde{z}^7}$\,(the black dash-dotted line).} \label{uz_rt0_combined}
\end{figure}

\begin{figure}
	\centerline{\includegraphics[scale=1.05]{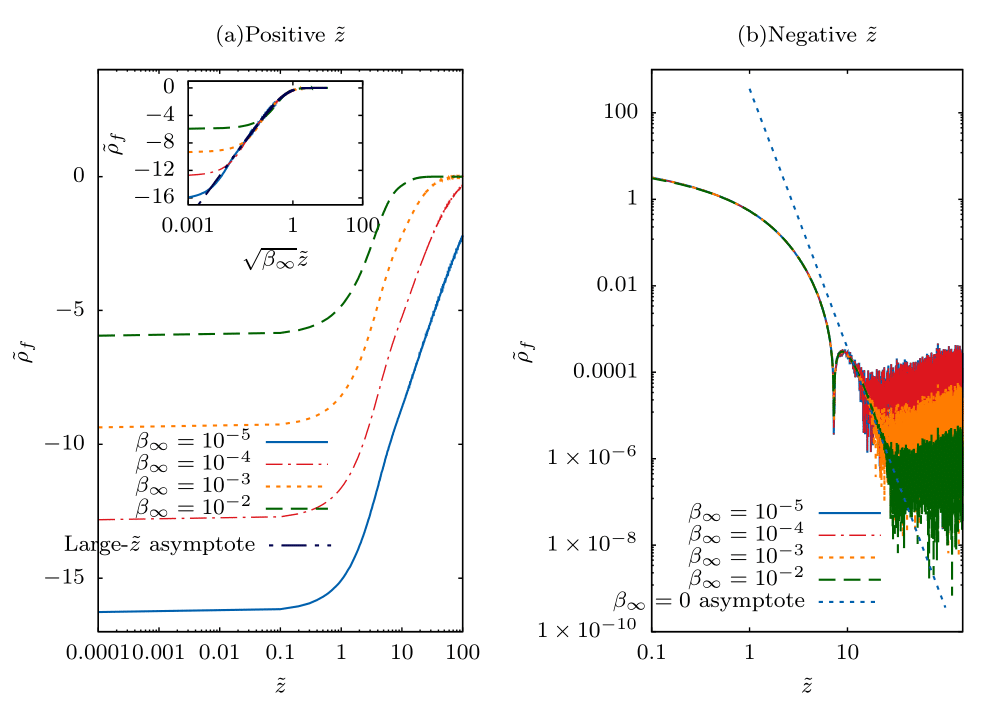}}% Images in 100% size
	\caption{The density disturbance field plotted along the stagnation streamline for both positive\,(LHS plot) and negative $\tilde{z}$\,(the RHS plot), and for different small $\beta_\infty$. The inset plot in the LHS figure shows the collapse of the density disturance profiles onto a common far-field asymptote, given by (\ref{eqn:axialfarjet2}), when plotted as a function of $\beta_\infty^{\frac{1}{2}}\tilde{z}$. The RHS plot shows that the small-$\beta_\infty$ density profiles converging to a common limiting form given by $\frac{360}{\tilde{z}^6}$; although in agreement with the farfield asymptote, the numerical approximations\,(with $N=1,50,000$) break down for large axial distances, with this breakdown being delayed the most for $\beta_\infty = 10^{-2}$.} \label{rhof_rt0_combined}
\end{figure}

Figures \ref{uz_rt0_combined} and \ref{rhof_rt0_combined} show plots of the axial velocity and density disturbance fields evaluated numerically at points along the stagnation streamline, based on (\ref{eqn:axvel2}) and (\ref{eqn:densdis2}), with $\bar{r}_t = 0$. In figure \ref{uz_rt0_combined}, the right hand side plot shows the transition of the axial velocity field, for negative $\tilde{z}$, from an $O(1/\tilde{z})$ Stokeslet decay in the inner region, to the more rapid $O(1/\tilde{z}^7)$ decay of the large-$\eta$ asymptote derived earlier\,(see section \ref{farfieldwake:largePe}), on length scales greater than the (primary)\,screening length. Note that this transition is accompanied by a reversal in direction, as evident from the sharp dip around $|z| \approx 8.85$ in the aforementioned logarithmic plot. Thus, the axial flow in the neighborhood of the front stagnation streamline, and at distances larger than the screening length, points towards the sphere. Importantly, the axial velocity profiles are virtually coincident for $\beta_\infty \leq 10^{-2}$, implying that the flow pattern in the vicinity of the front stagnation streamline converges to a limiting form for $Pe \rightarrow \infty$ that is characterized by the primary screening length of $O(Ri_v^{-\frac{1}{3}})$. In contrast, the plot on the left hand side, for positive $\tilde{z}$, shows a transition from the inner region Stokeslet decay to an eventual exponential decay at the largest distances, with this transition being postponed to progressively larger $\tilde{z}$ with decreasing $\beta_\infty$. For the smallest $\beta_\infty's\,(=10^{-4}$ and $10^{-5})$, one can see the emergence of an intermediate asymptotic regime, corresponding to $1 \ll \tilde{z} \ll \beta_\infty^{-\frac{1}{2}}$, where the velocity conforms to the reverse-Stokeslet behavior predicted above. Note that both the Stokeslet and reverse-Stokeslet behavior appear as the same asymptote\,(the black dot-dashed line), since the plot is for the absolute value of the velocity field on a logarithmic scale, and the indication of the reversal in direction is again the intervening sharp dip corresponding to $\tilde{z} \approx 1.15\,(z \approx 1.15Ri_v^{-\frac{1}{3}}$). The inset in this plot shows that the axial velocity profiles collapse onto a universal exponential behavior, when the ordinate and abscissa are rescaled with $\beta_\infty^{\frac{1}{2}}$ and $\beta_\infty^{-\frac{1}{2}}$, respectively, the latter corresponding to the axial distance being scaled by the secondary screening length. This collapse is consistent with (\ref{eqn:axialfarjet1}) above; although, since the distance corresonding to the reversal in direction scales with the primary screening length, the dips of the curves in the inset plot, are no longer coincident for varying $\beta_\infty$. 

The plots in figure \ref{rhof_rt0_combined} again highlight the contrast between the density disturbance fields along the front and rear stagnation streamlines. The plot on the right hand side, for negative $\tilde{z}$, shows that the density disturbance converges to a limiting form for $\beta_\infty \leq 10^{-2}$, with an $O(1/\tilde{z}^6)$ far-field decay, consistent with the large$-\eta$ asymptote obtained in section \ref{farfieldwake:largePe}; although, the numerics break down beyond a critical $|\tilde z|$ that is a function of the number of quadrature points used. In contrast, the left hand side plot shows that the density disturbance transitions from a near-field plateau to a far-field exponential decay, with this plateau increasing in magnitude logarithmically with decreasing $\beta_\infty$, consistent with (\ref{eqn:axialfarjet2}), precluding a collapse of the density profiles for small $\beta_\infty$. The inset in this figure plots the density profiles as a function of the rescaled abscissa, $\beta_\infty^{\frac{1}{2}}\tilde{z}$, so as to highlight their collapse onto a common curve\,(the modified Bessel function asymptote given by (\ref{eqn:axialfarjet2})). The individual curves deviating from this common asymptote on account of the near-field plateauing behavior, with this deviation occurring at a progressively smaller distance with decreasing $\beta_\infty$; note that for $\beta_\infty \rightarrow 0$, the said plateau regime becomes vanishingly small, while the exponential decay is pushed off to infinitely large distances\,(in units of the primary screening length), so the density field becomes logarithmically singular all along the rear stagnation streamline.

\subsubsection{Comparison of numerical profiles with the far-field approximations: Transition from the jet to the wake regimes}

\begin{figure}
	\centerline{\includegraphics[scale=1.1]{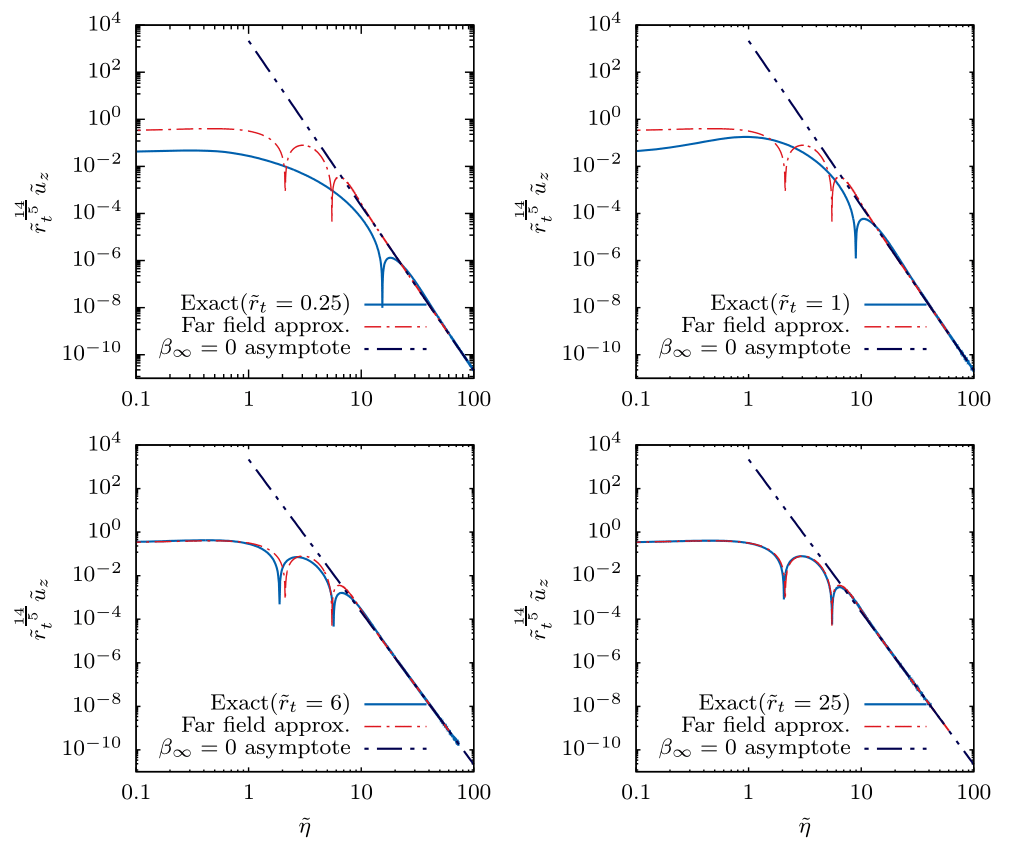}}% Images in 100% size
	\caption{The comparison, for negative $\tilde z$, between the numerically evaluated axial velocity profile, and the far-field wake-approximation given by (\ref{hpffweq2}), in the convection dominant limit, and in the Stokes stratification regime\,($Re \ll Ri_v^{\frac{1}{3}}$); the exact profile is obtained from a numerical integration of (\ref{eqn:axvel2}) with $\beta_\infty = 10^{-5}$.}
	\label{fig:hpfullfarneg}
\end{figure}

\begin{figure}
	\centerline{\includegraphics[scale=1.2]{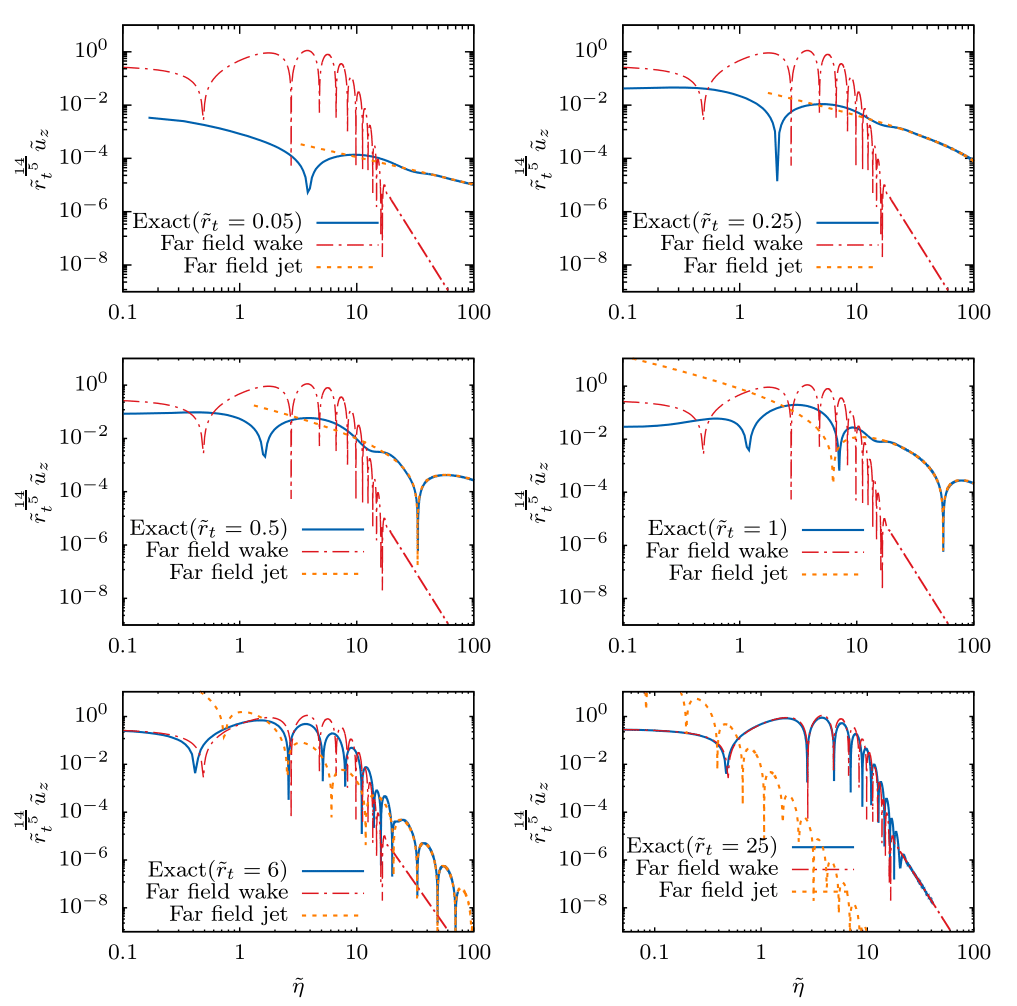}}% Images in 100% size
	\caption{The comparison, for positive $\tilde z$, between the numerically evaluated axial velocity profile, and both the far-field jet and wake-approximations given by (\ref{eqn:axialfarjet1}) and (\ref{hpffweq2}), respectively. The profiles pertain to the convection dominant limit and the Stokes stratification regime\,($Re \ll Ri_v^{\frac{1}{3}}$); the numerical profile is obtained from an integration with $\beta_\infty = 10^{-5}$.}
	\label{fig:hpfullfarpos}
\end{figure}

Having characterized the far-field approximations for the disturbance fields in both the buoyant jet\,(section \ref{jet:largePe}) and wake\,(section \ref{farfieldwake:largePe}) regions, we now compare the exact results for the axial velocity, obtained from a numerical evaluation of (\ref{eqn:axvel2}), with $\beta_\infty = 10^{-5}$,  with these approximations. The comparison is shown in Figures \ref{fig:hpfullfarneg} and \ref{fig:hpfullfarpos} for negative and positive $\tilde \eta$, respectively. Motivated by the self-similar one-dimensional integral approximation given by (\ref{hpffweq2}), both figures plot $\bar{r}_t^{\frac{14}{5}}|\tilde{u}_z|$ as a function of $\tilde{\eta}$. Only the wake-similarity profile (\ref{hpffweq2}) is relevant for negative $\tilde{\eta}$, and is shown alongside the exact numerical profiles in figure \ref{fig:hpfullfarneg} for different $\bar{r}_t$, together with its large-$\eta$ asymptotic form given by $2160/\tilde{\eta}^7$. The comparison here is similar to the diffusion dominant case, the agreement being poor for small to order unity $\bar{r}_t$, with the number of zero crossings also being in disagreement, but improving with increasing $\bar{r}_t$. There is good agreement for $\bar{r}_t = 6$, and almost a perfect match between the analytical and numerical profiles for $\bar{r}_t = 25$.

The comparison for positive $\tilde{\eta}$ is more elaborate since one now has both far-field wake and jet approximations in different regions of the half-space. One expects the axial velocity profile to transition from a jet-like profile to a wake-like one as one moves away from the rear stagnation streamline, that is, for a fixed $\tilde{z}$ and with increasing $\bar{r}_t$. This is seen in figure \ref{fig:hpfullfarpos} where the numerically determined axial velocity profiles are shown for six different $\bar{r}_t$'s ranging from $0.05$ to $25$, together with the far-field wake and jet approximations developed in the earlier two subsections. For the smallest $\bar{r}_t\,(=0.05)$, the exact calculation matches the far field jet approximation for $\tilde{z}$ greater than about $10$; for the chosen $\beta_\infty$, this jet approximation is virtually identical\,(in magnitude) to a Stokeslet decay over the range of $\tilde{\eta}$ examined. For the aforementioned $\bar{r}_t$, similar to figure \ref{uz_rt0_combined}, the numerical profile has a zero-crossing at a smaller $\tilde{z} \approx 1.15$, and continues to diverge at smaller $\tilde{z}$, in accordance with the expected Stokeslet behavior in the inner region, with there being the beginning of a plateau at the smallest $\tilde{z}$'s. For $\bar{r}_t = 0.25$, the plateauing tendency for small $\tilde z$ is more readily evident, with there still being a good agreement with the jet approximation for large $\tilde z$. The plateauing behavior arises for any finite $\bar{r}_t$ since the disturbance velocity field is now finite in the plane $\tilde{z} = 0$; the continued divergence down to $\tilde{z} = 0$ only occurs along the rear stagnation streamline\,(see figure \ref{uz_rt0_combined}). For $\tilde{r}_t$ values greater than unity, the exact profile starts to agree better with with the wake approximation, and for $r_t=25$ this agreement is near-perfect, with the jet approximation being virtually irrelevant. Although not shown, an analogous scenario prevails for the density disturbance profiles. 

From figures \ref{uz_rt0_combined} and \ref{fig:hpfullfarpos}, one sees that although the axial profile velocity exhibits only a single zero crossing along the rear stagnation streamline\,(corresponding to the Stokeslet-reverse-Stokeslet transition for $\beta_\infty = 0$), the jet-approximation for any non-zero $\tilde{r}_t$\,(the expression (\ref{jet:1Dintegral1})) appears to exhibit a denumerably infinite number of zero crossings as evident from the plots in the former figure for $\tilde{r}_t = 6$ and $\tilde{r}_t = 25$. The infinitely many zero-crossings suggest an infinite number of recirculating cells in the region $\tilde{z} \gg \tilde{r}_t$, $\tilde{z},\tilde{r}_t \gg 1$. Note that this conclusion is not necessarily in conflict with the wake approximation, that has only a finite number of zero-crossings, since the latter approximation is restricted to the region $\tilde{z} \ll \tilde{r}_t$. Thus, although the self-similar profiles in the wake predict an eventual algebraic decay, in reality, this decay might not extend to indefinitely large distances, but instead with increasing $\tilde{z}$, one will again have zero-crossings in the region $\tilde{z} > \tilde{r}_t$. As of now, this is difficult to verify, given the near-impossibility of accurate numerical evaluation at such large distances. Nevertheless, and although not evident from figures \ref{fig:hpstream_beta105} and \ref{fig:hpffwsim}, the implication of the above argument is that the flow-field in the convection-dominant limit exhibits an infinite number of recirculating cells\,(unlike the diffusion-dominant limit).

\begin{figure}
	\centerline{\includegraphics[width=1.25\linewidth,height=1.0\textheight]{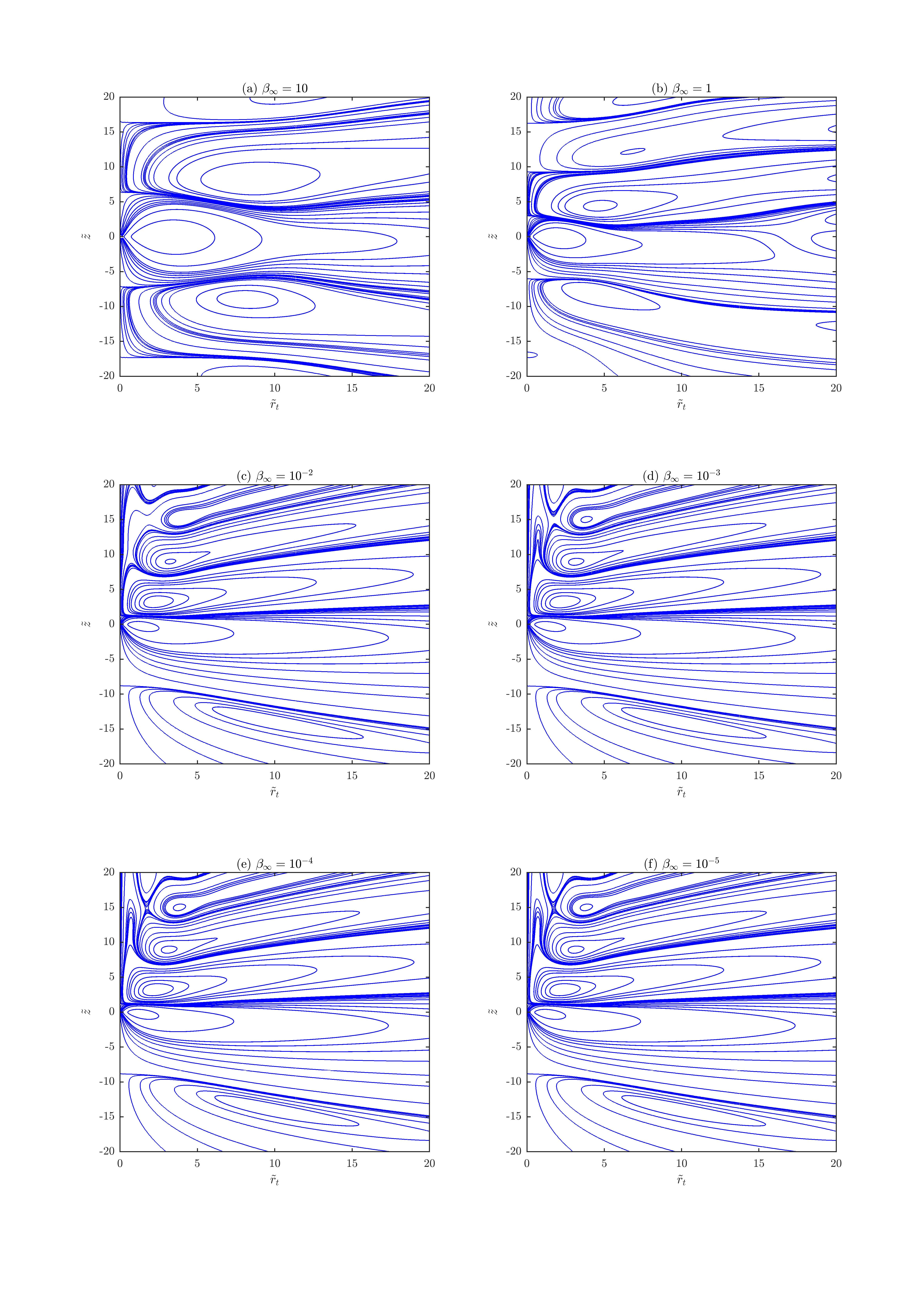}}% Images in 100% size
	\vspace*{-0.5in}
	\caption{Streamline patterns, pertaining to the Stokes-stratification regime\,(defined by the stratification screening length being the smallest of all relevant scales), for various $\beta_\infty$. The first plot for $\beta_\infty = 10$ is in the diffusion-dominant limit and nearly fore-aft symmetric; the plot for $\beta_\infty = 10^{-5}$ shows the buoyant reverse jet in the rear.}	\label{fig:hpstream}
\end{figure}

\begin{figure}
	\centerline{\includegraphics[scale=0.78]{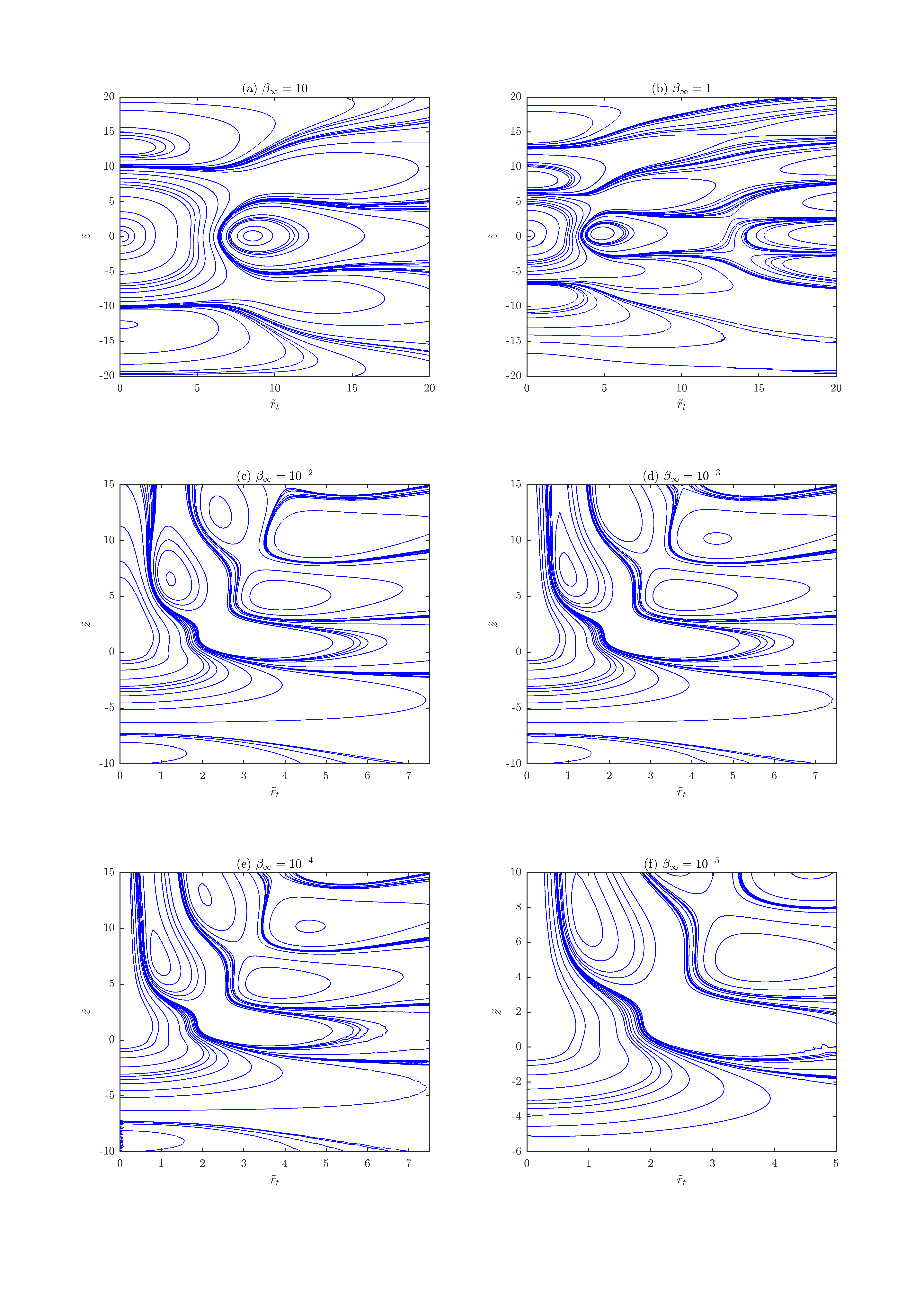}}% Images in 100% size
		\vspace*{-0.5in}
	\caption{Isopycnals pertaining to the Stokes-stratification regime for various $\beta_\infty$. The first plot for $\beta_\infty = 10$ is in the diffusion-dominant limit and nearly fore-aft symmetric; the plots for the smallest $\beta_\infty$'s are suggestive of a developing singularity along the rear stagnation streamline.}
	\label{fig:hpdensdist}
\end{figure}

Finally, figures \ref{fig:hpstream} and \ref{fig:hpdensdist} show the streamline and iso-pycnal patterns, respectively, for $\beta_\infty$ varying over the range $10^{-5}-10$. The departure of both patterns from fore-aft symmetry, with decreasing $\beta_\infty$, is evident, with the buoyant jet clearly evident in the streamline patterns for $\beta_\infty \leq 10^{-2}$. The spatial extent of all the streamline patterns shown corresponds to $\tilde{z}|,|\tilde{r}| \leq 20$, with these intervals measured in units of $Ri_v^{-\frac{1}{3}}$. For $\beta_\infty = 10^{-5}$, this implies that the streamline pattern includes the first two zero crossings that appear in the large-$\tilde{r}_t$ axial velocity profile in figure \ref{fig:hpfullfarpos}, while including both the zero crossings that appear in the profiles in figure \ref{fig:hpfullfarneg}. Note that the length scale characterizing the pattern changes from $Ri_v^{-\frac{1}{3}}$ to $(Ri_vPe)^{-\frac{1}{4}}$ with increasing $\beta_\infty$. In units of $Ri_v^{-\frac{1}{3}}$, this corresponds to the characteristic length scale increasing as $\beta_\infty^{\frac{1}{4}}$. Thus, for the same range in $\tilde{z}$ and $\tilde{r}_t$, one samples a proportionately smaller region of the streamline pattern with increasing $\beta_\infty$. This reduction in the spatial extent is evident from a comparison of the streamline pattern for $\beta_\infty = 10$ to the one in figure \ref{fig:lpcombined}. As seen in figure \ref{fig:hpdensdist}, the iso-pycnals become heavily compressed and distorted for the smallest $\beta_\infty$'s, in a manner consistent with the density disturbance having a logarithmic singularity along the rear stagnation streamline\,($\bar{r}_t = 0, \tilde{z} >0$) and as a result, numerically resolving the iso-pycnal becomes very difficult; this difficulty is reflected in the range of accessible $\tilde{r}_t$ and $\tilde{z}$ in figure \ref{fig:hpdensdist} progressively decreasing with decreasing $\beta_\infty$\,(this isn't an issue for the streamline patterns, given that the axial velocity remains finite along the rear stagnation streamline even for $\beta_\infty = 0$).

\subsection{Effects of weak inertia or convection} \label{secondary_inertia_stst}

In our calculations thus far, we have completely neglected the role of inertia. This is equivalent to assuming the inertial screening length\,(of $O(Re^{-1})$) to be much larger than the relevant stratification screening length, the latter being $(Ri_vPe)^{-\frac{1}{4}}$ for $Pe \ll 1$ and $O(Ri_v^{-\frac{1}{3}})$ for $Pe \gg 1$, with this ordering of the screening lengths corresponding to the Stokes stratification regime.  With regard to the calculations above, this is equivalent to setting $\alpha_0=0$ in (\ref{eq:vel}) and $\alpha_\infty=0$ in (\ref{oregion:largePe2}), for small and large $Pe$, respectively. While the detailed calculation of the flow field in the presence of competing effects of inertia and buoyancy is beyond the scope of the present manuscript, the effect of weak inertia on the larger-scale structure of the velocity field may nevertheless be inferred via simple scaling arguments. 

\begin{figure}\hspace*{-0.3in}
	\centerline{\includegraphics[scale=0.43]{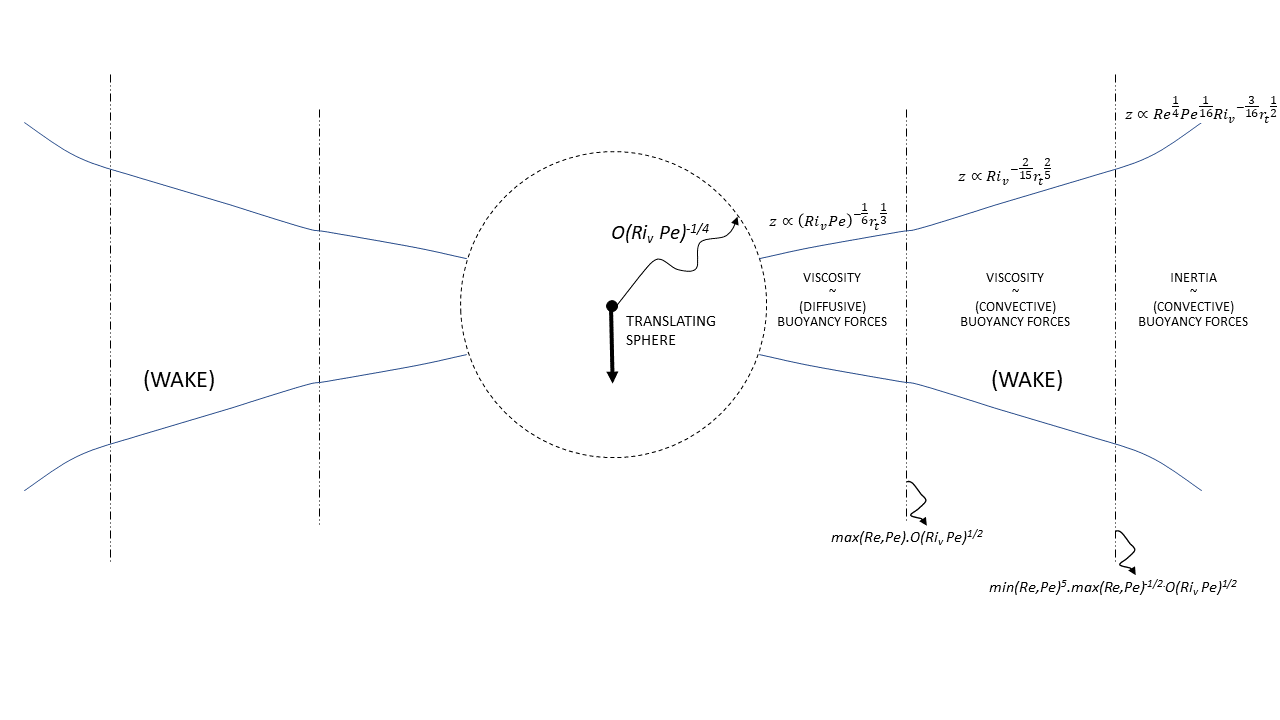}}% Images in 100% size
	\vspace*{-0.5in}
	\caption{Schematic of the different wake-growth regimes in the diffusion-dominant limit\,($Pe \ll 1$).}
	\label{fig:lpschema}
\end{figure}

We begin with the diffusion-dominant case, corresponding to $Pe \ll 1$ where, for small but finite $\alpha_0$, the denominator of the Fourier integrals for the disturbance fields, obtained from Fourier transforming  (\ref{eq:cont_Pe0})-(\ref{eq:dens_Pe0}), takes the form $\alpha_0\beta_0k_3^2k^2+\mathrm{i}k_3(\alpha_0+\beta_0)k^4+k^6+k_t^2$, with $k$ here being scaled by $(Ri_vPe)^{\frac{1}{4}}$. Note that the term proportional to $\beta_0k_3k^4$ denotes effects arising from the (weak)\,convection of the density disturbance field, and is typically associated with a screening length of $O(Pe^{-1})$ \citep{leal_2007}; the fore-aft asymmetry in the far-field arising from this term alone was already seen in the streamline and iso-pycnal patterns corresponding to the largest $\beta_\infty$'s in figures \ref{fig:hpstream} and \ref{fig:hpdensdist}. Assuming buoyancy forces to first become important with increasing distance from the settling sphere, we now know from section \ref{diffdom_results} that the dominant motion is restricted to an axisymmetric wake on distances greater than $O(Ri_vPe)^{-\frac{1}{4}}$, and is primarily horizontal. Thus, in order to examine inertia-induced transitions in the wake-scaling at larger distances, one may set $k_3 \gg k_t$, whence the aforementioned Fourier-space expression takes the form $\alpha_0\beta_0k_3^4+i(\alpha_0+\beta_0)k_3^5+k_3^6+k_t^2$. For $\alpha_0 = \beta_0 = 0$, one obtains the balance $k_3^6 \approx k_t^2$, and the vertical extent of the aforesaid wake growing as $z \propto (Ri_vPe)^{-\frac{1}{6}}r_t^{\frac{1}{3}}$\,(in units of $a$), as shown in section \ref{diffdom_results}. For $\alpha_0, \beta_0$ small but finite, the neglected terms invariably become important, and balance buoyancy forces\,(instead of viscosity) on larger lengthscales, corresponding to smaller $k$'s. For $\alpha_0 \ll \beta_0$\,(or $Re \ll Pe$), one obtains the balance $\beta_0 k_3^5 \approx k_t^2$ beyond a radial length scale of $O(Ri_v^{\frac{1}{2}}Pe^{-\frac{5}{2}})$; this balance is the same as that in section \ref{farfieldwake:largePe}, and therefore, implies a wake that grows as $z \propto Ri_v^{-\frac{2}{15}}r_t^{\frac{2}{5}}$. Thus, even for $Pe \ll 1$, one obtains the large-$Pe$ wake-scaling derived in section \ref{farfieldwake:largePe}, but only beyond the aforementioned secondary screening length. Finally, on the largest scales, the leading order balance is between inertial and buoyancy forces, and takes the form $\alpha_0\beta_0 k_3^4 \approx k_t^2$, leading to a growth of $z \propto Re^{\frac{1}{4}}Pe^{\frac{1}{16}}Ri_v^{-\frac{3}{16}} r_t^{\frac{1}{2}}$ beyond a radial scale of $Re^{-\frac{5}{2}}Ri_v^{\frac{1}{2}}$ that may be termed a tertiary screening length, again for $Re \ll Pe$. Thus, in the diffusion-dominant limit, weak convection\,(small but finite $Pe$) and inertia effects\,(small but finite $Re$) alter the far-field wake-scaling, causing it grow progressively faster beyond the screening lengths obtained above. Although the difference in the growth exponents is marginal\,($1/3 \rightarrow 2/5 \rightarrow 1/2$), one expects a more significant alteration of the wake structure; the change in structure accompanying the first transition in growth exponent\,($1/3 \rightarrow 2/5$) involves a departure from fore-aft symmetry, and the details may already be inferred from sections \ref{farfield:lowPe} and \ref{farfieldwake:largePe}. Provided the stratification screening length, $(Ri_vPe)^{-\frac{1}{4}}$, remains the smallest of the three possible primary screening lengths viz. $(Ri_vPe)^{-\frac{1}{4}}$, $Re^{-1}$ and $Pe^{-1}$, an assumption that defines the Stokes stratification regime for small $Pe$, the screening lengths derived above remain well ordered under the assumption $Re \ll Pe$. If we allow for convection and inertial effects to be small but of comparable importance, so that $\alpha_0/\beta_0 \sim O(1)$\,(or $Re \sim Pe$), then the growth exponents found above remain the same, but the secondary and tertiary screening lengths are now given by $max(Re,Pe)^{-3}(Ri_vPe)^{\frac{1}{2}}$ and $(min(Re,Pe)^{5}max(Re,Pe))^{-\frac{1}{2}}(Ri_vPe)^{\frac{1}{2}}$. A schematic of the different wake-scaling regimes in the diffusion-dominant limit is given in figure~\ref{fig:lpschema}; fluid motion outside the wake remains negligibly small.

\begin{figure}\hspace*{-0.3in}
	\centerline{\includegraphics[scale=0.43]{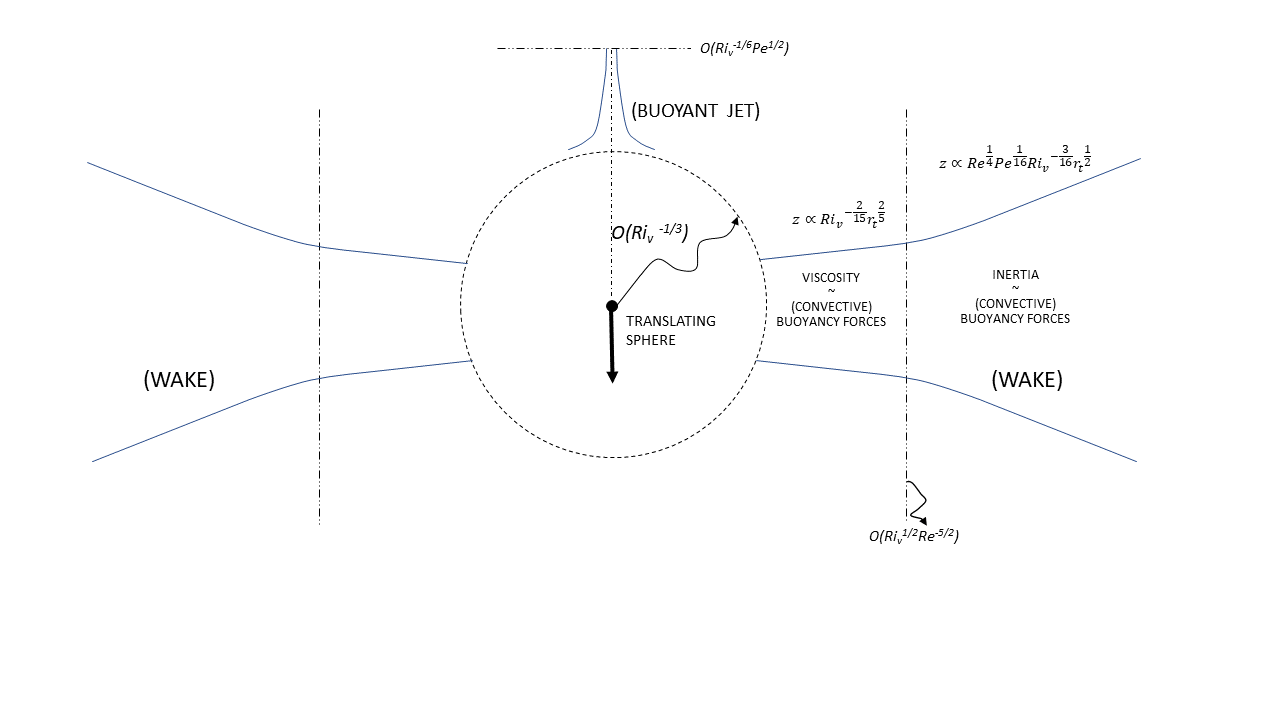}}% Images in 100% size
	\vspace*{-0.5in}
	\caption{Schematic of various regions in the convection dominant limit for non-zero Reynolds and Peclet numbers in the Stokes-stratification regime}
	\label{fig:hpschema}
\end{figure}

The effects of inertia in the convection dominant limit, corresponding to $Pe \gg 1$, is based on the same expression as that in the preceding paragraph, except that $k$ is now scaled with $Ri_v^{\frac{1}{3}}$, and accordingly, one has the form $-\alpha_\infty k_3^2 k^2+ i \alpha_\infty \beta_\infty k_3k^4+\mathrm{i}k_3k^4+\beta_\infty k^6+k_t^2$. Outside of the buoyant jet, one may neglect $\beta_\infty k^6$ and use $k_3 \gg k_t$ implying the dominance of horizontal motion. Setting $\alpha_\infty = \beta_\infty = 0$ then leads to the balance $k_3^5 \approx k_t^2$ which, as already seen in section \ref{farfieldwake:largePe}, and again in the analysis of the diffusion-dominant case above, yields the wake scaling $z \approx Ri_v^{-\frac{2}{15}} r_t^{\frac{2}{5}}$. At length scales larger than a radial threshold of $O(Re^{-\frac{5}{2}}Ri_v^{\frac{1}{2}})$, the balance is between the inertial and buoyancy forces, leading to the same square-root scaling $z \propto r_t^{\frac{1}{2}}$ seen above. Thus, the only difference with regard to the wake-scalings, in relation to the diffusion-dominant limit analyzed above, is that the initial scaling regime $z \propto r_t^{\frac{1}{3}}$ is now absent, and one directly transitions to the $z \propto r_t^{\frac{2}{5}}$ scaling regime at distances much greater than the stratification screening length of $O(Ri_v^{-\frac{1}{3}})$. As already mentioned in section \ref{jet:largePe}, a novel feature in the large-$Pe$ regime is the emergence of a buoyant jet that is smeared out by diffusion beyond a length scale of $O(Ri_v^{-\frac{1}{6}} Pe^{\frac{1}{2}})$. Again, provided the stratification screening length of $O(Ri_v^{-\frac{1}{3}})$ remains the smallest of the primary screening lengths, the diffusion-screening length for the jet is asymptotically smaller than the secondary screening length, of $O(Re^{-\frac{5}{2}}Ri_v^{\frac{1}{2}})$ above. A schematic of the various scaling regimes, in the convection-dominant limit, is given in figure~\ref{fig:hpschema}.

While the discussion in this manuscript has been restricted to the Stokes stratification regime, we briefly mention the screening lengths relevant to the inertia-stratification regime that, for large $Pe$, is defined by $Ri_v^{\frac{1}{3}} \ll Re$, or $\alpha_\infty \gg 1$; the inertial screening length of $O(Re^{-1})$ is now the primary screening length. For $Re \ll 1$, the fore-aft symmetric flow field in the inner Stokesian region first transitions, on scales of $O(Re^{-1})$, to a far-field structure consisting of an $O(1/r^2)$ source flow everywhere except for a viscous wake behind the translating sphere that acts as a directed sink\,\citep{batchelor_1967,subramanian_2010}. In terms of the Fourier-space expression given in the preceding paragraph, the viscous wake corresponds to the balance $\mathrm{i}k_3k^4 \sim \alpha_\infty k_3^2 k^2$, leading to the familiar scaling $r_t \sim (z/Re)^{\frac{1}{2}}$ for the wake growth in physical space. This source-wake structure is expected to be modified by buoyancy forces when $k_t^2$ becomes comparable to the terms in the aforementioned balance. This happens for $k \sim O(\alpha_\infty^{-\frac{1}{2}})$, which gives a secondary screening length of $O(Re^{\frac{1}{2}}Ri_v^{-\frac{1}{2}})$ in the inertia-stratification regime\,\citep{zhang_core_2019}. The structure of the flow field on these length scales is currently under investigation.

\section{Conclusions}\label{conclusions}
\subsection{Summary of main results}

We have analyzed in detail both the disturbance velocity and density fields induced by a sphere translating in a linearly stratified ambient fluid otherwise at rest. The analysis pertains to the Stokes stratification regime when buoyancy forces are dominant over inertial ones, so the transition from the well known Stokesian behavior, in the inner region, first occurs across a screening length determined by a balance between viscous and buoyancy forces. While we analyze the fluid motion in the diffusion-dominant limit\,(section \ref{diffdom_results}), this scenario has also been the focus of earlier work\,\citep{list_laminar_1971} and \citep{ardekani_2010}), and our main focus is therefore on the convection dominant limit\,($Pe \gg 1$) when the screening length is $Ri_v^{-\frac{1}{3}}$. In the latter limit, and within the Stokes stratification regime defined by $Re \ll Ri_v^{\frac{1}{3}} \ll 1$, we show through both numerical integration\,(section \ref{convecdom_results}) and asymptotic analysis\,(section \ref{farfieldwake:largePe}), that the far-field fluid motion consists of an axisymmetric wake surrounding the sphere whose vertical extent grows as $z \propto  Ri_v^{-\frac{2}{15}}r_t^{\frac{2}{5}}$, and wherein the fluid motion is predominantly horizontal; an analog of this wake also exists in the diffuson dominant limit, in which case it grows as $z \propto (Ri_vPe)^{-\frac{1}{6}}r_t^{\frac{1}{3}}$; $z$ and $r_t$ here being scaled by $a$. Although not obvious from the figures in earlier sections,the amplitude of fluid motion at a given non-dimensional distance (measured in units of the relevant screening length) is significantly greater for $Pe \gg 1$. In sharp contrast to the diffusion dominant limit, we have shown\,(section \ref{jet:largePe}) that there also exists a buoyant reverse jet in the vicinity of the rear stagnation streamline for $Pe \gg 1 $. Unlike the usual laminar or turbulent jets which broaden with increasing distance downstream on account of the momentum flux being conserved, the buoyant jet region above narrows down with increasing distance downstream as $r_t \propto Ri_v^{-\frac{1}{6}} z^{-\frac{1}{2}}$, with a velocity field that, although oppositely directed, decays in the same manner as a Stokeslet for $Pe = \infty$; the jet is screened by diffusion beyond a length scale of $O(Ri_v^{-\frac{1}{6}} Pe^{\frac{1}{2}})$ for large but finite $Pe$. The recent effort of \citet{shaik_2020b} has investigated the flow pattern due to a particle settling in the aforementioned convection dominant limit, based on a numerical fast Fourier transform technique. Although the primary emphasis was on calculating the drift volume, their examination of the fluid motion shows the existence of a strong reverse flow along the rear stagnation streamline, consistent with our findings. Finally, in section \ref{secondary_inertia_stst}, we comment briefly on the role of weak inertial (and convection) effects on the structure of the fluid motion beyond the primary buoyancy-induced screening length.

The fore-aft asymmetry of the large-$Pe$ disturbance velocity field found here has implications for pair-interactions. A vertically oriented particle-pair will experience a repulsive interaction for sufficiently large separations\,(on scales of $O(Ri_v^{-\frac{1}{3}})$). This is in contrast to the Stokesian scenario where the particle-pair separation remains invariant with time, a fact that may be established using reversibility arguments, and may be seen explicitly from the fore-aft symmetry of the Stokesian velocity field; note that the fore-aft symmetry of the $Pe = 0$ velocity field, obtained in section \ref{diffdom_results}, implies that the particle-pair separation, in a stratified fluid, is conserved to leading order in the diffusion dominant limit. For $Pe \gg 1$, the aforementioned repulsive pair-interaction is initially controlled by the greater magnitude of the velocity field along the front stagnation streamline, this because the zero-crossing along the front stagnation streamline\,($\approx 8.85Ri_v^{-\frac{1}{3}}$)  occurs at a greater distance than that on the rear stagnation streamline\,($\approx 1.15Ri_v^{-\frac{1}{3}}$). However, for distances a little beyond approximately $2 Ri_v^{-\frac{1}{3}}$, the more rapid $O(1/z^7)$ decay of the disturbance velocity in front of the particle implies that the repulsion is controlled by the slowly decaying $O(1/z)$ disturbance along the rear stagnation streamline. Succinctly put, the rear particle pushes the one in front for smaller separations, while the opposite is true at larger separations. The range of repulsion is limited to a length scale of $O(Ri_v^{-\frac{1}{6}} Pe^{\frac{1}{2}})$ by the effects of diffusion. This repulsive behavior is the opposite of the drafting behavior known for homogeneous fluids at finite $Re$.

\subsection{Discussion: the inner-region scaling estimates} \label{Inner_region}

It was indicated in the introduction as to how the validity of a linearized approximation is not obvious at large $Pe$, given that the ambient iso-pycnals in the inner region are severely distorted by the sphere velocity field. An examination of the density disturbance in the inner region for large $Pe$ should help identify possible restrictions on the results obtained in the manuscript, and a few comments in this regard are in order. We begin with the simpler case of small $Pe$ when the density perturbation around the sphere, on length scales of $O(a)$ (the inner region), remains finite at all times. The no-flux condition on the sphere surface causes the ambient iso-pycnals to tilt, so as to meet the sphere in a normal orientation. This tilting effect is significant in a region of $O(a^3)$, implying that the associated density perturbation is $O(\gamma a)$. The resulting baroclinically induced vorticity drives a flow of $O(\gamma a^3g/\mu)$, or $O(Ri_v)$ in non-dimensional terms (scaled by $U$; see \cite{stratified_torque_2021}). For $Ri_v \ll 1$, this weak flow may evidently be neglected compared to the primary Stokesian field. On larger length scales, convection of the base-state stratification by the perturbation Stokeslet field leads to a density perturbation that grows as $O(Pe\,r)$ in the inner region. The buoyancy forcing due to this density perturbation becomes comparable to viscous forces on length scales of $O(Ri_vPe)^{-\frac{1}{4}}$, the small-$Pe$ stratification screening length screening length identified first by \cite{list_laminar_1971} and \cite{ardekani_2010}. Importantly, for small $Pe$, the Stokesian flow remains a valid leading order approximation in the inner region for all times.

For large $Pe$, the density perturbation  in the inner region can become much larger than the nominal $O(\gamma a)$ estimate above. This may be seen by considering the limiting case of $Pe = \infty$, when the iso-pycnals are affinely convected by the sphere velocity field. The sphere, as it settles through the stably stratified medium, entrains positively buoyant fluid in a quasi-spherical annular region that extends behind it in a narrow wake that lengthens with time. The amplitude of the density perturbation near the sphere increases linearly with time as $O(\gamma Ut)$, leading to a buoyancy forcing per unit volume of $O(\gamma Ut g)$. Clearly, for large enough times, this buoyancy forcing will become comparable to the viscous terms even in the inner region, and for $Ri_v \ll 1$. Since the viscous terms in the equations of motion are $O(\frac{\mu U}{a^2})$ in the inner region, the threshold time at which buoyancy forces are of a comparable magnitude is $O(\frac{\mu}{\gamma a^2 g})$, or $O(\frac{a}{U} Ri_v^{-1})$. This is therefore the time at which the flow in the inner region must deviate from the leading Stokesian approximation on account of buoyancy forces; as mentioned in the introduction, it is still possible for the structure of the fluid motion to remain similar to that detailed in this manuscript, but for a buoyancy-induced renormalization of the force exerted by the particle, although only a detailed examination of the inner region would confirm this. Moving to the outer region, in the Stokes stratification regime, the time scale associated with the development of the flow field in this region may be estimated as the time required for momemtum to diffuse to a distance of $O(aRi_v^{-1/3})$, which is $O(\frac{a^2}{\nu} Ri_v^{-2/3})$. The ratio of this latter time to the time scale estimated above, for the inner region to depart from a homogeneous Stokesian evolution, is $O(Re Ri_v^{1/3})$, and therefore, asymptotically small for $Re$, $Ri_v \ll 1$. Thus, there is an asymptotically long interval of time corresponding to $\frac{a^2}{\nu}Ri_v^{-2/3} \ll t \ll \frac{a}{U} Ri_v^{-1}$, where one has a quasi-steady response in the outer region, with the motion in the inner region still governed by the Stokes equations at leading order. The findings with regard to the nature of the fluid motion, detailed in section \ref{convecdom_results}, are certainly valid in this time period. Note that for any finite $Pe$, however large, the distortion of the isopycnals will not continue indefinitely. Instead, there will eventually be a steady state boundary layer, of thickness $O(aPe^{-\frac{1}{3}})$, as far as the density gradient is concerned\,(although not for the density itself which will continue to increase with time for an assumed constant $U$).

Scaling arguments similar to those in the preceding paragraph may also be used to assess the possibility of observing quasi-steady dynamics on scales beyond the primary screening length, and thereby, examine the relevance of the wake-scaling regimes sketched in section \ref{secondary_inertia_stst}; see figures \ref{fig:lpschema} and \ref{fig:hpschema}. Focusing on the Stokes stratification regime for large $Pe$,  the arguments in section \ref{secondary_inertia_stst} pointed to a secondary screening length of $O(aRe^{-\frac{5}{2}}Ri_v^{\frac{1}{2}})$ across which the dominant balance shifted from one between buoyancy and viscous forces to one between buoyancy and inertial forces. Given that the inertial forces enter the dominant balance, the time scale for a quasi-steady wake to be established on the aforementioned secondary screening length may be estimated as $\frac{a Re^{-\frac{5}{2}}Ri_v^{\frac{1}{2}}}{U}$. The ratio of this time scale to $\frac{aRi_v^{-1}}{U}$ gives us $Re^{-\frac{5}{2}}Ri_v^{\frac{3}{2}}$, with this ratio needing to be much less than unity in order for a quasi-steady analysis of the fluid motion to hold; this yields $Re \gg Ri_v^{\frac{3}{5}}$. Combining this with the primary criterion for the large-$Pe$ Stokes stratification regime gives $Ri_v^{\frac{3}{5}} \ll Re \ll Ri_v^{\frac{1}{3}}$ for the dynamics in both the primary and secondary outer regions to be quasi-steady, in the time that the inner region region has a Stokesian character.

\subsection{Discussion: the drift volume scaling estimates} \label{Drift_estimates}

We now turn to the drift volume estimate for a sphere settling in a density-stratified fluid which, as mentioned in the introduction, was one of the motivations for the analysis in this paper. The rapid algebraic decay of the far-field velocity disturbance, induced by buoyancy forces, implies that the drift volume\,(${\mathcal D}$) will be finite in presence of an ambient stratification, as originally argued by \citet{subramanian_2010}. More precise estimates for ${\mathcal D}$ as a function of $Ri_v$ and $Re$, in the Stokes and inertia-stratification regimes, are obtained below. For the homogeneous Stokesian scenario, the $O(1/r)$ decay of the disturbance field implies a divergent drift volume for any finite time. As originally shown by \citet{eames_2003}, it therefore becomes necessary to define a partial drift volume\,(${\mathcal D}_p$) where, in contrast to \citet{darwin_note_1953}, one only considers an initial material plane of a finite spatial extent. In a recent effort, \citet{chisholm_2017} have shown that, at leading order in $a/h$, ${\mathcal D}_p \sim ah^2 \sinh^{-1} (Ut/h)$, $t$ and $h$ here being the time and radius of the aforementioned material plane, respectively; the $h$-scaling clearly points to the finite-time divergence of ${\mathcal D}\,(= \lim_{h \rightarrow \infty} {\mathcal D}_p$) in the Stokesian limit. In the limits $Ut/h \ll 1$ and $Ut/h \gg 1$, the authors find ${\mathcal D}_p$ to be $O(ahUt)$ and $O[ah^2\ln (Ut/h)]$, respectively. These scalings may be readily obtained without a detailed calculation: for $Ut \ll h$, the flux through the original plane is independent of time, and due to the $U$-component of the Stokesian field, in the transverse plane containing the sphere. This component is $3Ua/(4r_t)$, and the flux through a circular section of radius $h$ is therefore given by $\textstyle\int_0^h 3U(a/4r_t) 2\pi r_t dr_t \approx (3\pi /2)Uah$, implying ${\mathcal D}_p \approx (3\pi/2)Uah t$; here, the lower limit of the integral is taken to be $0$ since the leading contribution comes from $r_t$ of $O(h)$\,(this is also the reason why a Stokeslet approximation suffices for the leading order estimate). In the long-time limit of interest, when the distance of the material plane from the sphere is much larger than its radial extent, the flux is primarily due to the velocity $u_z \approx 3U/az$ along the rear stagnation streamline. The drift displacement due to this disturbance velocity field may be estimated as  $\textstyle\int^t dt\, u_z = \textstyle\int^{Ut} (dz'/U)u_z = \textstyle\int^{Ut} dz'(3a/2z') \sim  (3a/2)\ln (Ut)$, and is logarithmically divergent in time. A subtle point here is with regard to the argument of the logarithm; the approximate estimate above gives a dimensional argument for the logarithm, and one needs an additional length with respect to which $Ut$ in the logarithm is measured. Although an obvious choice would be $a$, the correct choice is $h$\,(as also evident from the exact result above), and this is because the onset of the logarithmic divergence is dependent on the transverse radial location of the fluid element. The decreasing magnitude of the disturbance field implies that it takes a progressively longer time for an element, further off from the translation axis, to be displaced through a distance of $O(a)$; evidently, the logarithmic divergence in time can only begin after the drift displacement has attained a magnitude of $O(a)$. For an element at a transverse distance of $O(h)$, the scales that contribute dominantly to ${\mathcal D}_p$, this time is $O(h/U)$, implying that the argument of the logarithm, in the expression for the drift displacement above, should be $t/(h/U)$; multiplication by $\pi h^2$ gives the estimate ${\mathcal D}_p \approx \frac{3\pi}{2} ah^2\ln (Ut/h)$. 

In the Stokes-stratification regime, one expects the dominant contribution to the drift volume to come from the range $h \sim l_c$, $l_c$ being the relevant stratification screening length; $l_c \sim O[(Ri_vPe)^{-\frac{1}{4}}]$ for $Pe \ll 1$, and $O(Ri_v^{-\frac{1}{3}})$ for $Pe \gg 1$. However, for elements at these distances\,(from the translation axis), the drift displacement attains a magnitude of $O(a)$ only in the $O(l_c/U)$ time taken for the sphere to translate through a screening length. Since the velocity field decays faster for larger separations, there cannot be the analog of the aforementioned logarithmic-in-time behavior, for larger times, that occurred in the homogeneous case. This implies that ${\mathcal D}$ for the Stokes drift displacement can be obtained from the aforementioned long-time estimate for the Stokesian case by replacing $h$ with $l_c$, but removing the logarithm. One therefore obtains ${\mathcal D} \sim O[a^3(Ri_vPe)^{-\frac{1}{2}}]$ and $O(a^3 Ri_v^{-\frac{2}{3}}$), for small and large $Pe$, in the Stokes stratification regime, the latter estimate being relevant to the oceanic scenario\,\citep{katija_2009,subramanian_2010}; both estimates diverge in the limit of a homogeneous ambient\,($Ri_v \rightarrow 0$), as must be the case. The numerical pre-factors in these estimates would require a detailed calculation of the drift displacements on length scales of order the stratification screening length. Note that fluid elements that start off at distances of $h \ll l_c$ from the translation axis will suffer drift displacements of $O(a \ln Ri_v^{-\frac{1}{3}})$, and one therefore expects higher-order terms involving logarithmics in a small $Ri_v$ expansion of ${\mathcal D}$ in the limit $Re \ll Ri_v^{\frac{1}{3}} \ll 1$. Recent efforts by \citet{shaik_2020} and \citet{shaik_2020b} have obtained ${\mathcal D}_p$ numerically, in both the small and large $Pe$ limits, Consistent with the results of \citet{chisholm_2017}, ${\mathcal D}_p$ exhibits an $O(h^2)$ scaling with the radial extent of the material plane under consideration. Importantly, however, the scaling arguments above imply that this algebraic divergence must be cut off once $h \sim O(l_c)$. A more detailed examination of pathlines and drift volume calculation to support the scaling arguments in this paragraph will be reported in a separate communication.

In the inertia-stratification regime\,($Ri_v \ll Re \ll 1$), discussed briefly towards the end of section \ref{secondary_inertia_stst}, the disturbance velocity field attains the familiar source-sink structure on length scales larger than the primary\,(inertial) screening length of $O(aRe^{-1})$\,\citep{batchelor_1967}. It is well known that the presence of a viscous wake leads to ${\mathcal D}$ diverging linearly in time for the homogeneous scenario\,\citep{subramanian_2010,chisholm_2017}. This divergence is readily seen from the constant flux through a fixed plane driven by the viscous wake. This flux is given by $u_z (r_t^{wake})^2$, where $r_t^{wake}\sim (az/Re)^{\frac{1}{2}}$, and is $O(Ua^2/Re)$, leading to ${\mathcal D} \sim (Ua^2/Re)t$ for the homogeneous case. For the stratified case, and for $Pe \gg 1$, this viscous wake only persists until the secondary screening length of $O(Re/Ri_v)^{-\frac{1}{2}}$ obtained in section \ref{secondary_inertia_stst}, and therefore the linear divergence above will be cut off for $t \sim O(\frac{a(Re/Ri_v)^{-\frac{1}{2}}}{U})$, when stratification forces screen the wake velocity field, and one obtains ${\mathcal D} \sim a^3 (ReRi_v)^{-\frac{1}{2}}$ in the inertia-stratification regime. Note that this scaling is consistent with the scaling obtained above in the Stokes-stratification regime, in that it reduces to $O(Ri_v^{-2/3})$ for $Re=Ri_v^{1/3}$. In summary, for a fixed $Ri_v \ll 1$, ${\mathcal D}$ starts off being $O(a^3 Ri_v^{-\frac{2}{3}})$ until an $Re of O(Ri_v^{\frac{1}{3}})$, decreasing thereafter as $O(a^3Re^{-\frac{1}{2}} Ri_v^{-\frac{1}{2}})$ for $Re \gg Ri_v^{\frac{1}{3}}$.

\nocite{*}
\section*{Acknowledgements}
Numerical computations reported here are carried out using the Param Yukti facility provided under the National Supercomputing Mission, and the Nalanda-2 computational cluster available with JNCASR. The authors thank the institute for providing these facilities.
\appendix
\section{The far-field wake velocity and density fields in the diffusion-dominant\,($Pe = 0$) limit}\label{appendix1}

Herein, we start with the expressions (\ref{fflpeq1}) and (\ref{fflpeq2}), where the approximation of nearly horizontal motion has already been made. In a cylindrical coordinate system aligned with the translation direction, and after carrying out the $\phi$ integration, the expressions for the axial and transverse velocities, and the density disturbance, reduce to:
\begin{equation}
\bar{u}_{z}=\frac{-3}{2\pi}\int_{-\infty}^\infty dk_3 \int_0^\infty dk_t \frac{k_3^2k_t^3 J_0(k_t\bar{r}_t)e^{ik_3\bar{z}}}{(k_3^6+k_t^2)},
\end{equation}

\begin{equation}
\bar{u}_{r_t}=\frac{3i}{2\pi}\int_{-\infty}^\infty dk_3 \int_0^\infty dk_t \frac{k_3^3k_t^2 J_1(k_t\bar{r}_t)e^{ik_3\bar{z}}}{(k_3^6+k_t^2)},
\end{equation}

\begin{equation}
\bar{\rho}_{f_1}=\frac{-3}{2\pi}\int_{-\infty}^{\infty} dk_3 \int_0^\infty dk_t \frac{k_t^3 J_0(k_t\bar{r}_t)e^{ik_3\bar{z}}}{(k_3^6+k_t^2)}
\end{equation}

Next, one uses contour integration to evaluate the $k_3$-integral. Contributions arise from the existence of six poles in the complex-$k_3$ plane, with these poles being symmetrically disposed about the real $k_3$-axis, consistent with the fore-aft symmetry of the disturbance fields. The contour integration yields the following one-dimensional integrals:

\begin{equation}
\bar{u}_{z}=-3i\int_0^\infty k_t^2 J_0(k_t \bar{r}_t)\left(lq_1^2e^{i q_1 k_t^{\frac{1}{3}} \abs{\bar{z}}}+mq_2^2e^{i q_2 k_t^{\frac{1}{3}} \abs{\bar{z}}}+nq_3^2 e^{i q_3 k_t^{\frac{1}{3}} \abs{\bar{z}}}\right)dp, 
\end{equation}
\begin{equation}
\bar{u}_{{r_t}}=-3\sign{\bar{z}} \int_0^\infty k_t^{\frac{4}{3}} J_1(k_t \bar{r}_t) \left(lq_1^3 e^{i q_1 k_t^{\frac{1}{3}} \abs{\bar{z}}}+mq_2^3 e^{i q_2 k_t^{\frac{1}{3}} \abs{\bar{z}}}+nq_3^3 e^{i q_3 k_t^{\frac{1}{3}} \abs{\bar{z}}}\right)dp, 
\end{equation}
\begin{equation}
\bar{\rho}_{f}=-3i\int_0^\infty k_t^{\frac{4}{3}} J_0(k_t \bar{r}_t) \left(l e^{i q_1 k_t^{\frac{1}{3}} \abs{\bar{z}}}+m e^{i q_2 k_t^{\frac{1}{3}} \abs{\bar{z}}}+n e^{i q_3 k_t^{\frac{1}{3}} \abs{\bar{z}}}\right)dp.
\end{equation}
Here $q_1$, $q_2$, $q_3$, $l$, $m$ and $n$ are complex-valued constants given by:
\begin{align}
[q_1,q_2,q_3,q_4,q_5,q_6]=[e^{\frac{\pi i}{6}}, e^{\frac{\pi i}{2}},e^{\frac{5 \pi i}{6}}, e^{\frac{7 \pi i}{6}}, e^{\frac{9 \pi i}{6}},e^{\frac{11 \pi i}{6}}], \nonumber \\
l=\frac{1}{(q_1-q_2)(q_1-q_3)(q_1-q_4)(q_1-q_5)(q_1-q_6)}, \nonumber \\
m= \frac{1}{(q_2-q_1)(q_2-q_3)(q_2-q_4)(q_2-q_5)(q_2-q_6)}, \nonumber \\
n= \frac{1}{(q_3-q_1)(q_3-q_2)(q_3-q_4)(q_3-q_5)(q_3-q_6)}. \nonumber
\end{align}
Setting $k_t\bar{r}_t = p$ as the integration variable, and using $\eta=\frac{\bar{z}}{\bar{r_t}^{\frac{1}{3}}}$, with some simplification, yields (\ref{lpffweq1}),  (\ref{lpffweq3}) and (\ref{lpffweq2}).

\section{The far-field wake velocity and density fields in the convection-dominant\,($Pe \rightarrow \infty$ or $\beta_\infty \rightarrow 0$) limit}\label{appendix2}

Herein, we start with the expressions appropriate for the far-field wake, given by (\ref{fflpeq1}) and (\ref{fflpeq2}), where the approximation of nearly horizontal motion has already been made. In a cylindrical coordinate system aligned with the translation direction, and after carrying out the $\phi$ integration, the expressions for the axial and transverse velocities, and the density disturbance, reduce to:
\begin{equation}
\bar{u}_{z}=\frac{-3i}{2\pi}\int_{-\infty}^\infty dk_3 \int_0^\infty dk_t \frac{k_3k_t^3 J_0(k_t\bar{r}_t)e^{ik_3\bar{z}}}{(ik_3^5+k_t^2)}
\end{equation}

\begin{equation}
\bar{u}_{r_t}=\frac{3i}{2\pi}\int_{-\infty}^\infty dk_3 \int_0^\infty dk_t \frac{k_3^2k_t^2 J_1(k_t\bar{r}_t)e^{ik_3\bar{z}}}{(ik_3^5+k_t^2)}
\end{equation}

\begin{equation}
\bar{\rho}_{f_1}=\frac{-3}{2\pi}\int_{-\infty}^{\infty} dk_3 \int_0^\infty dk_t \frac{k_t^3 J_0(k_t\bar{r}_t)e^{ik_3\bar{z}}}{(ik_3^5+k_t^2)}
\end{equation}
As for the diffusion dominant case analyzed in appendix \ref{appendix1}, the next step is to evaluate the $k_3$-integral using contour integration. There now exist five poles in the complex-$k_3$ plane with two poles in the lower half and three poles in the upper half of the complex plane; the differing number of poles in the two halves of the plane translates to fore-aft asymmetry of the axial velocity and density disturbance fields. The residue integration then yields the following one dimensional integrals for positive and negative $\tilde{z}$:
\begin{align}
\tilde {u}_{z}&=-3i\int_0^\infty k_t^\frac{9}{5}J_0[k_tr_t][Q_1q_1e^{iq_1k_t^{\frac{2}{5}}\tilde{z}}+Q_2q_2e^{iq_2k_t^{\frac{2}{5}}\tilde{z}}+Q_3q_3e^{iq_3k_t^{\frac{2}{5}}\tilde{z}}]dk_t \textrm{ } (\textrm{for } \tilde z>0) \nonumber \\
&=3i\int_0^\infty k_t^\frac{9}{5}J_0[k_tr_t][Q_4q_4e^{iq_4k_t^{\frac{2}{5}}\tilde{z}}+Q_5q_5e^{iq_5k_t^{\frac{2}{5}}\tilde{z}}]dk_t \textrm{ } (\textrm{for } \tilde z<0)
\end{align}
\begin{align}
\tilde {u}_{r_t}&=-3\int_0^\infty k_t^\frac{6}{5}J_0[k_tr_t][Q_1q_1^2e^{iq_1k_t^{\frac{2}{5}}\tilde{z}}+Q_2q_2^2e^{iq_2k_t^{\frac{2}{5}}\tilde{z}}+Q_3q_3^2e^{iq_3k_t^{\frac{2}{5}}\tilde{z}}]dk_t \textrm{ } (\textrm{for } \tilde z>0) \nonumber \\
&=3\int_0^\infty k_t^\frac{6}{5}J_0[k_tr_t][Q_4q_4^2e^{iq_4k_t^{\frac{2}{5}}\tilde{z}}+Q_5q_5^2e^{iq_5k_t^{\frac{2}{5}}\tilde{z}}]dk_t \textrm{ } (\textrm{for } \tilde z<0)
\end{align}
\begin{align}
\tilde {\rho}_{f}&=-3\int_0^\infty k_t^\frac{7}{5}J_0[k_tr_t][Q_1e^{iq_1k_t^{\frac{2}{5}}\tilde{z}}+Q_2e^{iq_2k_t^{\frac{2}{5}}\tilde{z}}+Q_3e^{iq_3k_t^{\frac{2}{5}}\tilde{z}}]dk_t \textrm{ } (\textrm{for } \tilde z>0) \nonumber \\
&=3\int_0^\infty k_t^\frac{7}{5}J_0[k_tr_t][Q_4e^{iq_4k_t^{\frac{2}{5}}\tilde{z}}+Q_5e^{iq_5k_t^{\frac{2}{5}}\tilde{z}}]dk_t \textrm{ } (\textrm{for } \tilde z<0)
\end{align}

Here $q_1$, $q_2$, $q_3$, $q_4$, $q_5$, $Q_1$, $Q_2$, $Q_3$, $Q_4$ and $Q_5$ are complex-valued constants given by:
\begin{align}
[q_1,q_2,q_3,q_4,q_5]=[e^{\frac{\pi i}{10}}, e^{\frac{\pi i}{2}},e^{\frac{9 \pi i}{10}}, e^{-\frac{7 \pi i}{10}}, e^{-\frac{3 \pi i}{10}}], \nonumber \\
Q_1=\frac{1}{(q_1-q_2)(q_1-q_3)(q_1-q_4)(q_1-q_5)}, \nonumber \\
Q_2= \frac{1}{(q_2-q_1)(q_2-q_3)(q_2-q_4)(q_2-q_5)}, \nonumber \\
Q_3= \frac{1}{(q_3-q_1)(q_3-q_2)(q_3-q_4)(q_3-q_5)}, \nonumber \\
Q_4= \frac{1}{(q_4-q_1)(q_4-q_2)(q_4-q_3)(q_4-q_5)}, \nonumber \\
Q_5= \frac{1}{(q_5-q_1)(q_5-q_2)(q_5-q_3)(q_5-q_4)}. \nonumber
\end{align}
Setting $k_t\bar{r}_t = p$ as the integration variable, and using $\eta=\frac{\tilde{z}}{\tilde{r_t}^{\frac{2}{5}}}$ yields (\ref{hpffweq1}), (\ref{hpffweq2}) and (\ref{hpffweq3}).
%\section{Effects of inertia and convection in the limit of $Pe<<1$}\label{appA}

\bibliographystyle{jfm}
\bibliography{stratified_1_fields}
\end{document}